\documentclass{PoS}

\usepackage{amsmath}

\newcommand{\be}{\begin{equation}}
\newcommand{\ee}{\end{equation}}
\def\beqa{\begin{eqnarray}}
\def\eeqa{\end{eqnarray}}
\def\bean{\begin{eqnarray*}}
\def\eean{\end{eqnarray*}}
\def\nn{ }

\newcommand{\R}{\mathbb{R}}
\newcommand{\C}{\mathbb{C}}

\newcommand{\tg}{{\tilde{g}}}
\newcommand{\te}{{\tilde{e}}}

\newcommand{\tI}{{\tilde{I}}}
\newcommand{\tJ}{{\tilde{J}}}
\newcommand{\tX}{{\tilde{X}}}

\newcommand{\tA}{{\tilde{A}}}

\newcommand{\bI}{{\mathbf{I}}}
\newcommand{\bJ}{{\mathbf{J}}}

\newcommand{\bQ}{{\mathbf{Q}}}

\newcommand{\dd}{{\mathrm{d}}}

\newcommand{\eqn}[1]{(\ref{#1})}
\newcommand{\del}{\partial}
\newcommand{\Tr}[1]{\:{\rm Tr}\,#1}

\title{Principal Chiral Model without and with WZ term: Symmetries and Poisson-Lie T-Duality }

\ShortTitle{Principal Chiral Model}

\author{ \speaker{Francesco Bascone} \thanks{Talk ``Symmetries and Dualities in Sigma Models with Wess-Zumino term" delivered at the Workshop on Quantum Geometry, Field Theory and Gravity, September 18-25, 2019.}\\
       Dipartimento di Fisica ``E. Pancini'', Universit\`a di Napoli Federico II and  INFN-Sezione di Napoli, Complesso Universitario  di Monte S. Angelo Edificio 6, via Cintia, 80126 Napoli, Italy.\\
       E-mail: \email{francesco.bascone@na.infn.it}}

\author{ \speaker{Franco Pezzella} \thanks{Talk ``Principal Chiral Model: T-Duality Symmetries and Doubling" delivered at the Conference on Recent Developments in Strings and Gravity, September 10-16, 2019. }
\\
         INFN - sezione di Napoli, Complesso Universitario di Monte S. Angelo ed. 6, via Cintia - 80126 Napoli, Italy\\
        E-mail: \email{franco.pezzella@na.infn.it}}
        
\abstract{Duality properties of the $SU(2)$ Principal Chiral Model are investigated starting from a one-parameter family of its equivalent Hamiltonian descriptions generated 
by a non-Abelian deformation of the cotangent space $T^*SU(2) \simeq SU(2) \ltimes \mathbb{R}^3$. The corresponding dual models are obtained through $O(3,3)$ duality transformations  
and result to be defined on the group $SB(2,\mathbb{C})$, which is the Poisson-Lie dual of $SU(2)$ in the Iwasawa decomposition of the Drinfel'd double $SL(2,\mathbb{C})=SU(2) \bowtie SB(2,\mathbb{C})$.
These dual models provide an explicit realization of Poisson-Lie T-duality. A doubled generalized parent action is  then built on the tangent space $TSL(2,\mathbb{C})$. Furthermore, a generalization of the $SU(2)$ PCM  with a
 WZ term is shortly discussed.

        }

\FullConference{Corfu Summer Institute 2019 "School and Workshops on Elementary Particle Physics and Gravity" (CORFU2019)\\
		31 August - 25 September 2019\\
		Corfu, Greece}

\begin{document}

\section{Introduction}
\label{secintro}

T-duality provides a powerful tool for investigating the structure of the spacetime from the string point of view by relating, in the usual sigma-model approach, backgrounds which otherwise would be considered different \cite{GPR, AAL, D}. It sheds light on the short distance behavior of string theory, still  exotic at least from the perspective of quantum field theory. It is also a clear indication that ordinary geometric concepts can break down at the string scale. 

T-duality has been a valuable guide for finding new phenomena such as D-branes, mirror symmetry or exotic solutions. In the case of compactification of the space coordinate $x^{a}$ on a circle of radius $R$, T-duality is encoded, for bosonic closed strings, in the simultaneous transformations $R \leftrightarrow \alpha'/R$ and $p_{a}  \leftrightarrow w^{a} / \alpha'$  under which  the string coordinate $X^{a} = X^{a}_{L}+X^{a}_{R}$ is transformed in the dual coordinate $ \tilde{X}_{a} \equiv X^{a}_{L}-X^{a}_{R} $, being $p_{a}$ the quantized momentum along $x^{a}$ and $w^{a}$ the corresponding winding mode which plays the role of momentum for $\tilde{X}_{a}$.  More generally, T-duality allows to build new string backgrounds which could not be addressed otherwise and generally go under the name of  \textit{non-geometric backgrounds} (see e.g. \cite{Plauschinn2018} for a recent review on the subject).

On a $d$-torus $T^{d}$, with constant backgrounds provided by the tensor metric field $G_{a b}$ and the Kalb-Ramond field $B_{ab}$,  T-duality is described by $O(d,d; \mathbb{Z})$ transformations. By exchanging momentum and winding modes, it implies that the short distance behavior is governed by the long distance behavior in the dual torus $\tilde{T}^{d}$.

It has to be observed that, already at the classical level, the indefinite orthogonal group $O(d, d; \mathbb{R})$ naturally appears in the Hamiltonian description of the bosonic string with a world-sheet embedded into a $d$-dimensional target space $M$  together with two peculiar structures, the {\em generalized metric} ${\cal H}$ and the $O(d,d)$ invariant metric $\eta$ \cite{Rennecke}.

With $*$ the Hodge operator with respect to $h = \mbox{diag} (-1,1)$, the usual Polyakov world-sheet action for a bosonic string reads as:
\begin{eqnarray}
S[X; G, B] = \frac{T}{2} \int \left[ G_{ab}(X) (*dX^{a}) \wedge dX^{b} +
 B_{ab}(X) dX^{a} \wedge
dX^{b} \right] \nonumber 
\end{eqnarray}
with $G$ a Riemannian metric on the target space and $B$ a two-form.
Varying $S$ with respect to $X^{a}$ yields the equation of motion:
\begin{eqnarray}
d *dX^{a} + \Gamma^{a}_{\,\,bc} dX^{b} \wedge *dX^{c} = \frac{1}{2} G^{am}H_{mbc}dX^{b} \wedge dX^{c} \nonumber
\end{eqnarray}
with $H=dB$ and  $\Gamma^{a}_{\,\,bc} = \frac{1}{2} G^{am}(\partial_{b} G_{mc} + \partial_{c} G_{mb} - \partial_{m}G_{bc})$ the coefficients of the Levi-Civita connection.

The dynamics of the theory is determined by the equations of motion for the  coordinates $X^{a}$ accompanied with the constraints (in the conformal gauge):
\begin{eqnarray}
G_{ab} ( \dot{X}^{a}  \dot{X}^{b} +  X'^{a}  X'^{b} )=0  \,\,\,\,\,\,\,; \,\,\,\,\,\,\, G_{ab} \dot{X}^{a}  X'^{b} =0 \,\,\,\,\,\, \,\, \label{constraints}
\end{eqnarray}
deriving from the vanishing of the energy-momentum tensor $T_{\alpha \beta} \equiv \frac{\delta S}{\delta h^{\alpha \beta}}=0$, i.e. from the equation of motion for a general world-sheet metric $h$.

The Hamiltonian density $H$ is given by a Legendre transformation with respect to the canonical momentum $P_{a}=\frac{\partial L}{\partial \dot{X}^{a} }=\frac{1}{2 \pi \alpha'} \left( G_{ab} \dot{X}^{b} + B_{ab} X'^{b} \right)$ and $\dot{X}^{a}$, but results also from  a Legendre transformation with respect to the canonical winding ${W}_{a}= \frac{ \partial L}{ \partial X'^{a}} = - \frac{1}{2 \pi \alpha'} \left( G_{ab}  X'^{b} +B_{ab}  \dot{X}^{b} \right) $ and $X'^{a}$ since the first constraint  in eq. (\ref{constraints})  implies $P_{a}\dot{X}^{a}=W_{a}X'^{a}$.

The Hamiltonian density $H = \frac{1}{4 \pi \alpha'} G_{ab} ( \dot{X}^{a} \dot{X}^{b} + X'^{a}X'^{b}) $
 %equivalently 
can be rewritten as:
\begin{eqnarray}
 H  
 & = &
 \frac{1}{4 \pi \alpha'} 
\left( \begin{array}{c} \partial_{\sigma} X \nonumber \\
   2 \pi \alpha' P \end{array}\right)^{t} {\cal H}(G,B) \left( \begin{array}{c} \partial_{\sigma} X \nonumber \\ 2 \pi \alpha' P \end{array} \right) %\nonumber  
   \\
& = &  \frac{1}{4 \pi \alpha'} 
\left( \begin{array}{c} \partial_{\tau} X \nonumber \\
  - 2 \pi \alpha' W \end{array}\right)^{t} {\cal H}(G,B) \left( \begin{array}{c} \partial_{\tau} X \nonumber \\ -2 \pi \alpha' W \end{array} \right) 
\end{eqnarray}
where the {\em generalized metric} is introduced:
\begin{eqnarray}
{\cal  H}(G,B) \equiv \left(\begin{array}{cc}
           G-BG^{-1}B  & BG^{-1} \\
           -G^{-1}B & G^{-1}\end{array}\right)  \nonumber
\end{eqnarray}
{\em Generalized vectors }$A_{P}$ and $A_{W}$  can be  then defined 
in $TM \bigoplus T^{*}M$ as follows:
\begin{eqnarray}
A_{P}(X)  &=  &\partial_{\sigma} X^{a} \frac{\partial}{\partial x^{a}} + 2 \pi \alpha' P_{a} dx^{a}  \nonumber \\
A_{W}(X) & = & \partial_{\tau} X^{a} \frac{\partial}{\partial x^{a} }- 2 \pi \alpha' W_{a} dx^{a}
 \nonumber
\end{eqnarray}

Hence, one can see that $H$ results to be proportional to the squared length of the generalized vectors $A_{P}$  and $A_{W}$ 
as measured by the generalized metric ${\cal H}$.

% with components
%\begin{eqnarray}
%A_{P}(X) = \left( \begin{array}{c} \partial_{\sigma} X \nonumber \\ 2 \pi \alpha' P \end{array} \right) \,\,\,\,\,\,\,\,\,\,\,\,\,\,\,\,\,\,
%A_{W}(X) = \left( \begin{array}{c} \partial_{\tau} X \nonumber \\ 2 \pi \alpha' W \end{array} \right)  
%\end{eqnarray}

In particular, in terms of  $A_{P}$ the constraints in eq. (\ref{constraints})
%\begin{eqnarray}
%G_{ab} ( \dot{X}^{a}  \dot{X}^{b} +  X'^{a}  X'^{b} )=0  \,\,\,\,\,\,\,; \,\,\,\,\,\,\, G_{ab} \dot{X}^{a}  X'^{b} =0  \nonumber
%\end{eqnarray}
 respectively become:     
 \begin{equation}
A_{P}^{t} {\cal H} A_{P}=0     \,\,\,\,\,\,\,\,\,\, A_{P}^{t} {\cal \eta} A_{P}=0 . \nonumber
\end{equation}
 The first sets $H$ to zero while the second completely determines the dynamics,  rewritten in terms of the $O(d,d)$ invariant metric:
 $ {\cal \eta}  = \left(\begin{array}{cc}
           0  & {\bf 1} \\
          {\bf 1} & 0\end{array}\right)$. %i.e.  to the group $O(D,D)$.  
         This group is defined by the $d \times d$ matrices ${\cal T}$ satisfying the condition ${\cal T}^{t} {\cal \eta} {\cal T} = {\cal \eta}$.
 The generalized metric itself is an element of $O(d,d; \mathbb{R})$, since it satisfies: $ {\cal \eta} {\cal H} {\cal \eta} = {\cal H}^{-1}$, i.e. ${\cal H}^{t} {\cal \eta} {\cal H} = {\cal \eta}$.

All the admissible generalized vectors satisfying the constraint $A_{P}^{t} {\cal \eta} A_{P}=0$ are related by an $O(d,d; \mathbb{R})$ transformation via $A'_{P}={\cal T} A_{P}$.
Then, for $A'_{P}$ to solve the first constraint, a compensating transformation ${\cal T}^{-1}$ has to be applied to ${\cal H}$. , i.e. ${\cal H'} = ({\cal T}^{-1})^{t} {\cal H} ( {\cal T}^{-1} )$.

${\cal H}$ and its inverse ${\cal H}^{-1}$ can be rewritten in products:
\begin{eqnarray}
    {\cal  H}(G,B) =             \left(\begin{array}{cc}    
            1 & B \\         
            0 & 1 \end{array} \right) 
            \left(\begin{array}{cc}    
            G & 0 \\         
            0 & G^{-1} \end{array} \right)  
            \left(\begin{array}{cc}    
            1 & 0 \\         
            -B & 1 \end{array} \right)   \nonumber 
            \end{eqnarray}
            \begin{eqnarray}
   {\cal  H}^{-1}(G,B) =             \left(\begin{array}{cc}    
            1 & 0 \\         
            B & 1 \end{array} \right) 
            \left(\begin{array}{cc}    
            G^{-1} & 0 \\         
            0 & G \end{array} \right)  
            \left(\begin{array}{cc}    
            1 & -B \\         
            0 & 1 \end{array} \right)          \nonumber 
  \end{eqnarray}
 This indeed shows that the background $B$ can be created from the $G$-background through a transformation involving $B$, hence named $B$-transformation. Further examples of $B$-transformations will be introduced later when the Principal Chiral Model will be discussed.
    
In the case of compactification on  the $d$-torus  $T^{d}$ $O(d,d;\mathbb{R}) \rightarrow O(d,d;\mathbb{Z})$.

On $T^{d}$ , where $G_{ab}$ and $B_{ab}$ constant,  with its isometries $U(1)^{d}$, the e.o.m.'s  for the string coordinates are a set of conservation laws on the world-sheet \cite{Maharana}:
\begin{eqnarray}
\partial_{\alpha} J^{\alpha}_{a}=0 \,\,\,\,\,\,\,\, \mbox{with} \,\,\,\,\,\,\,\,J^{\alpha}_{a}=h^{\alpha \beta} G_{ab} \partial_{\beta} X^{b} + \epsilon^{\alpha \beta} B_{ab} \partial_{\beta} X^{b}  h \equiv  \epsilon^{\alpha \beta} \partial_{\beta} \tilde{X}_{a}  \,\, . \nonumber %\,\,\,\,\, \rightarrow \mbox{ dual coordinates}      \nonumber
\end{eqnarray}
%Locally, they can be expressed as:
%\begin{eqnarray}
 %h^{\alpha \beta} G_{ab} \partial_{\beta}X^{b} + \epsilon^{\alpha \beta} B_{ab}\partial_{\beta} X^{b}  \equiv  \epsilon^{\alpha \beta} \partial_{\beta} \tilde{X}_{a}   \,\,\,\,\, %\rightarrow \mbox{ dual coordinates}    \nonumber
%\end{eqnarray}

%\pause

Through the use of auxiliary fields, one gets the dual Polyakov action $\tilde{S}[\tilde{X}; \tilde{G}, \tilde{B}] $ on $\tilde{T}^{d}$ written in terms of the dual string coordinates $\tilde{X}_{a}$ and connected to $S[X; G, B]$ by $X^{a} \rightarrow \tilde{X_{a}}$ and suitable transformations of  $(G,B) \rightarrow  (\tilde{G}, \tilde{B}$) through the so-called B\"uscher rules \cite{Buscher1987, Buscher1988}. More specifically, 
in terms of the coordinates $\tilde{X}_{a}$'s, one can introduce 
 the action involving auxiliary fields $U_{\alpha}^{a}$ \cite{Maharana}:
\begin{eqnarray}
\! \! \! S[U;G,B] \! \!  = \!\!\! \int d^{2} \sigma \left\{ \left[ h^{\alpha \beta} U_{\alpha}^{a} U_{\beta}^{b} G_{ab} + \epsilon^{\alpha \beta} U_{\alpha}^{a} U_{\beta}^{b} B_{ab} \right] + \epsilon^{\alpha \beta} \partial_{\alpha} \tilde{X}_{a} U^{a}_{\beta} \right\}  \nonumber 
\end{eqnarray}
Varying with respect to $\tilde{X}_{a}$ gives  $\epsilon^{\alpha \beta} \partial_{\alpha} U_{\beta}^{a}=0 $, while the $U_{\alpha}^{a}$ equation of motion is:
\begin{eqnarray}
h^{\alpha \beta} U_{\beta}^{b} G_{ab}+ \epsilon^{\alpha \beta} U_{\beta}^{b} B_{ab} - \epsilon^{\alpha \beta} \partial_{\beta} \tilde{X}_{a} = 0.   \nonumber
\end{eqnarray}
This can be used to solve for $U_{a}^{\alpha}$ yielding:
\begin{eqnarray}
U_{\alpha}^{a} = \left( \epsilon_{\alpha}^{\,\,\, \beta} \tilde{G}^{a b} + \delta_{\alpha}^{\beta} \tilde{B}^{ab} \right) \partial_{\beta} \tilde{X}_{b}  \nonumber
\end{eqnarray}
\begin{eqnarray}
\tilde{S}[\tilde{X}; \tilde{G}, \tilde{B}] = \frac{T}{2} \int \left[ \tilde{G}_{ab} d\tilde{X}^{a} \wedge *d\tilde{X}^{b} + \tilde{B}_{ab}(X) d\tilde{X}^{a} \wedge
d \tilde{X}^{b} \right] \nonumber 
\end{eqnarray}
 being $\tilde{G} = (G-BG^{-1}B)^{-1}$ and $\tilde{B}=-G^{-1}B\tilde{G}$.

%\pause
In this case one refers to  {\em Abelian T-duality} for stressing the presence of global Abelian isometries in the target spaces of both the paired sigma models. 

For closed strings, toroidal compactification on $T^{d}$ implies the following periodicity conditions:
\begin{eqnarray}
X^{a} (\sigma, \tau) \equiv X^{a} (\sigma + \pi, \tau) + 2 \pi L^{a}  \,\,\,\,\,\,
L^{a} = \sum_{i=1}^{d} w^{i} R_{i} e_{i}^{a}   \nonumber
\end{eqnarray}
with $w^{i}$ being the winding numbers and $e_{i}^{a}$ vector basis on $T^{d}$.
In this case, $O(d,d; \mathbb{R})$ gets into $O(d,d;\mathbb{Z})$
and this constitutes the T-duality group of the toroidal compactification that provides a  symmetry not only of the mass
spectrum and the vacuum partition function but also of the scattering amplitudes. 

It results very natural  to require the string world-sheet sigma model  to be, therefore,  
manifestly $O(d,d)$ invariant (we will omit, in the following, the specification  $\mathbb{R}$ or $\mathbb{Z}$) \cite{D, Hull2005, Tseytlin1990, Tseytlin1991, Siegel1993, Lee2014}.  Such formulation is based on a {\em doubling} of the string coordinates along the compact directions of the target space, i.e. on the introduction of both the usual coordinates $X^{a}$ and their respective duals $\tilde{X}_{a}$.
The equations of motion for the {\em doubled} coordinates $\mathbb{X}^{A}  \equiv (X^{a}, \tilde{X}_{a})$ $(A=1, \dots 2d), (a=1, \dots d)$  can be combined into a single $O(d,d)$-invariant equation \cite{D}:
\begin{eqnarray}
{\cal H} \partial_{\alpha} \mathbb{X}  = {\cal \eta} \epsilon_{\alpha \beta} \partial^{\beta}  \mathbb{X}   \nonumber
\end{eqnarray}
For  $G_{ab} = \eta_{ab}$ and $B=0$, these reproduce the well-known Hodge duality conditions: $\partial_{\alpha} X^{a} =  \epsilon_{\alpha \beta} \partial^{\beta} \tilde{X}^{a}$.

%For closed string coordinates, the dual coordinates satisfy the same periodicity conditions and then the Abelian T-duality $O(d,d;\mathbb{Z})$  
%maps two theories of the same type into one another  providing a symmetry, not only of the mass
%spectrum and the vacuum partition function, but also of the scattering amplitudes. This symmetry is not manifest in the Polyakov action. 

After doubling the coordinates, i.e. putting the coordinates $X^{a}$  and the dual ones $\tilde{X}_{a}$ in the {\em generalized} vector $\mathbb{X}^{A}$ %= (X^{a}, \tilde{X}_{a})$ $(A=1, \dots 2d), (a=1, \dots d)$ 
above introduced,  it is natural to replace the standard world-sheet action of string theory based on $G$ and $B$ by an action written in terms of  $\eta$ and ${\cal H}$ and that could be manifestly invariant under Abelian T-duality. %to combine the equations of motion for these  coordinates %$\mathbb{X}^{A} = (X^{a}, \tilde{X}_{a})$ $(A=1, \dots 2d), (a=1, \dots d)$  can be combined 

The action proposed in ref. \cite{Tseytlin1991} fulfills this requirement and, for constant backgrounds, highlights the role of the generalized vector $\mathbb{X}$ and of the two metrics: 
\begin{eqnarray}
S=-\frac{T}{2} \int d\tau d\sigma \left[  \partial_{\tau}  \mathbb{X}^{A} \partial_{\sigma} \mathbb{X}^{B}  \eta_{AB} - \partial_{\sigma}  \mathbb{X}^{A} \partial_{\sigma}  \mathbb{X}^{B} {\cal H}_{AB}
 \right]   \label{Tseytlinact}
\end{eqnarray}
%with $\eta$ and $H$ constant.

This provides a  manifestly T-duality $O(d,d)$ symmetric formulation that may be considered as a natural generalization {\em at the string scale} of the usual Polyakov action that can be actually reproduced
at compactification radius $R >> \alpha'$ while its dual can be obtained at $R << \alpha'$. The new action, therefore, embodies 
the core  of T-duality on flat compact target spaces: the short distance behavior is governed by the long distance behavior in the dual space. One refers to it as the {\em doubled world-sheet action} \cite{Nibbelink2013, 
Angelis2014, Copland2012, Park2013, Berman2008, Ma2015, Berman2015, 
Pezzella2015, Pezzella2015a, Bandos2015}. 
From a manifestly T-dual invariant two-dimensional string world-sheet, Double Field Theory (DFT) \cite{HZ} should emerge out as a low-energy limit. DFT developed as a way to encompass the Abelian T-duality in field theory with Generalized and Doubled Geometry underlying it \cite{hitchin1, hitchin2, gualtieri:tesi}.

The doubled world-sheet action in  ref. (\ref{Tseytlinact}) can be also understood in terms of {\em Born geometry} that is based on the concept of Born reciprocity principle \cite{FRS}. Briefly, this latter states that the validity of Quantum Mechanics implies a fundamental symmetry between space and momentum that is broken by General Relativity because it is states that spacetime is curved, while energy-momentum space, i.e. the cotangent space, is linear and flat.
The simple but radical idea proposed by Max Born, is that in order to unify Quantum Mechanics and General Relativity one should also allow the phase space, and thus momentum space, to carry curvature. The action in (\ref{Tseytlinact}) is defined on the phase space with coordinates $\mathbb{X}^{A}  \equiv (X^{a}, \tilde{X}_{a})$ where the two metrics $\eta$ and ${\cal H}_{AB}$ have been introduced. When the corresponding sigma-model can be relaxed away from constant $\eta$ and ${\cal H}_{AB}$, this means that not only will space-time become curved, but momentum space as well. This then will lead to an implementation of Born reciprocity.

Beyond the Abelian T-duality, one could ask if there are similar aspects in more general settings, for
instance, when target space is curved or non-compact and what is the role played by the existence of group isometries. The answer to this question is provided by the introduction of two other kinds of T-dualities: non-Abelian T-duality and Poisson-Lie T-duality.
Non-Abelian T-duality refers to the existence of a global Abelian isometry on the target space of one of the two sigma models and of a global non-Abelian isometry on the other \cite{Ossa1993, AAL, RV93}.
In particular, in ref. \cite{Ossa1993}, the authors gauged the non-Abelian isometries of the sigma-models and constrained the field strength $F$ to vanish. The dual action was obtained by integrating out the gauge fields with the Lagrange multipliers becoming coordinates of the dual manifold.  The resulting non-Abelian dual turned out to lack the isometry that would make it possible to perform the duality transformation back to the original model. But, even without isometries, this dual is still equivalent to an apparently different sigma-model.
The notion of non-Abelian T-duality is still lacking some of the key features of its Abelian counterpart. The non-Abelian isometry group of the dual space was always smaller.
A canonical procedure is still missing that would yield the original theory if one is given its non-Abelian dual. 
The {\em Poisson-Lie T-duality} \cite{Klimcik1996a, Klimcik1996} generalizes the previous definition to all the other possible cases.  The relevant algebraic structure for the existence of a non-Abelian T-duality and a Poisson-Lie T-duality is not necessarily the existence of the group of isometries of the background, but some other structure that shows up only in special cases as an isometry group. 
%In fact, the relevant structure for the existence of  a dual counterpart  of a Principal Chiral Model on a Lie group $G$ lies on the  definition of Drinfel'd double together with the notion of Poisson-Lie symmetries.
%The Poisson-Lie T-duality relies therefore on the concept of Drinfel'd double when the target space is a group manifold.
In particular, a category of models that reveal themselves to be very helpful in understanding the above mentioned T-dualities is provided by sigma models whose target configuration space is a Lie group $G$ manifold. Particularly interesting among these models are the ones having as a target space  a Lie group $D$ whose Lie algebra ${\mathfrak d}$ can be decomposed into a pair of maximally isotropic subalgebras with respect to a non-degenerate invariant bilinear form on ${\mathfrak d}$. Such a structure is the Drinfel'd double \cite{Drinfeld1987}, in which the duality transformation simply exchanges the roles of the two groups forming the double. In this case it is said that the models are indeed dual in the sense of the Poisson-Lie T-Duality. We are going to analyze, from this perspective, the Principal Chiral Models (PCM) \cite{MPV2019, MPV2018} mostly without a Wess-Zumino term. Some final considerations will be made however also in the case of the presence of this latter \cite{BPV}. In fact, the Wess-Zumino-Witten model is involved in several interesting applications of string theory. In particular, it can describe strings propagating on a Lie group manifold, which are especially relevant backgrounds since they are Ricci flat and also represent the appropriate setting to analyze T-duality generalizations.  The WZW model on $SL(2,\mathbb{R})$ has been used to construct bosonic string theories on $AdS_3$ \cite{MO1, MO2, MO3}, or having $PSU(1,1|2)$ as target space it describes superstring theories on $AdS_3 \times S^3$ geometries \cite{gotz}. Another important application is provided by having a non semi-simple Lie group as target space, as the latter revealed to be particularly relevant as string backgrounds \cite{nappi, kehagias, wittenbh}.

 %a Lie group $D$ whose Lie algebra ${\mathfrak d}$ can be decomposed into a pair of maximally isotropic subalgebras with respect to a non-degenerate invariant bilinear form on ${\mathfrak d}$ . 
%\item
%Such a structure is the \textcolor{red}{Drinfel'd double}, in which the duality transformation simply exchanges the roles of the two groups forming the double. In this case it is said that the models are indeed dual in the sense of the {\em Poisson-Lie T-Duality}.
%\pause

\section{Principal Chiral Model}

A Principal Chiral Model is defined by  a sigma-model whose target space is a Lie group $G$. Studying this kind of models has led to abandoning the requirement of the existence of isometries for the target space as the condition for the existence of dual counterparts. Indeed, the relevant structure in this case reveals to be the one of Drinfel'd double for $G$ together with the well-established notion of Poisson-Lie symmetries \cite{Sfetsos:1998, Stern:1998, Stern:1999, Falceto:2001, Calvo:2003, Bonechi:2003, Sfetsos:2009, Severa:2017, Hassler:2017, Jurco:2017}.

 The Drinfel'd double of a Lie group $G$ is defined as a Lie group $D$, with dimension twice the one of $G$, such that its Lie algebra $\mathfrak{d}$ can be decomposed into a pair of maximally isotropic sub-algebras, $\mathfrak{g}, \tilde {\mathfrak{g}}$ with respect to a non-degenerate invariant bilinear form on $\mathfrak{d}$, with $ \mathfrak{g}, \tilde {\mathfrak{g}}$   respectively the Lie algebra of G and its dual algebra. The dual algebra is endowed with a Lie bracket which has to be compatible with existing structures.  %The relevant algebraic structure for the existence of a non-Abelian T-duality and a Poisson-Lie T-duality is not necessarily the existence of the group of isometries of the background, but some other structure that shows up only in special cases as an isometry group. 
%In fact, the relevant structure for the existence of  a dual counterpart  of a Principal Chiral Model on a Lie group $G$ lies on the  definition of Drinfel'd double together with the notion of Poisson-Lie symmetries.
%The Poisson-Lie T-duality relies therefore on the concept of Drinfel'd double when the target space is a group manifold.

%The Drinfel'd double  $D$ of the Lie group $G$ is a Lie group $D$, with  dimension twice the one of $G$. Its Lie algebra $\mathfrak{d}$ can be decomposed into a pair of maximally isotropic sub-algebras, $\mathfrak{g}, \tilde {\mathfrak{g}}$  with respect to a non-degenerate invariant bilinear form on $\mathfrak{d}$. %with $\mathfrak{g},   \tilde{\mathfrak{g}}$. 
Any such triple, $( \mathfrak{d}, \mathfrak{g},\tilde{\mathfrak{g}})$, is referred to as a Manin triple. By exponentiation of $\tilde{\mathfrak{g}}$ one gets the dual Lie group $\tilde{G}$  such that locally $D \simeq G \times \tilde{G}$. The simplest example is provided by the cotangent bundle of any $d$-dimensional Lie group $G$, $T^*G \simeq  G \ltimes \mathbb{R}^d$  which can be named {\em the classical double}, with trivial Lie bracket for the dual algebra  $ \tilde{\mathfrak{g}} \simeq \mathbb{R}^d$. 
For every  decomposition of the Drinfel'd double  $D$ into dually related  subgroups $G, \tilde{ G}$,  it is possible to define a couple of PCM's having as  target configuration space either of the two subgroups. 
 Every PCM has its dual counterpart for which the role of $G$ and its dual $\tilde{G}$ is interchanged : $ G \leftrightarrow \tilde{G}$. 
 The set of all decompositions $(\mathfrak{ d}, \mathfrak{ g},  \tilde{\mathfrak{g}}) $, plays the role of the modular space of sigma models mutually connected by an $O(d,d)$ transformation which turns to play the role of a duality transformation. 
The standard Abelian T-duality refers to the presence of Abelian isometries $U(1)^{d}$  in both the dual sigma models and they can be composed into $U(1)^{2d}$, the simplest example of a Drinfel'd double and  its modular space is  in one-to-one correspondence with $O(d,d)$ \cite{Klimcik1996}.

\subsection{SU(2) Principal Chiral Model}

Let us consider, in particular, an
$SU(2)$  Principal Chiral Model, that is to say a sigma model with target space $SU(2) $ and source space $\mathbb{R}^{1,1}$ endowed with the metric $h_{\alpha \beta}=\mathrm{diag}(1,-1)$. 

It can be shown that the target phase space of the $SU(2)$ PCM can actually be replaced by the Drinfel'd double of $SU(2)$, namely the group $SL(2, \mathbb{C})$, without modifying the dynamics \cite{R89,R892}.

This observation makes such model very interesting, since it allows to discuss Poisson-Lie T-duality in a framework in which it is a true symmetry of the undeformed dynamics.

\subsection{Lagrangian formalism}

In the Lagrangian approach  the action may be written in terms of fields $\phi : (t,\sigma) \in  \R^{1,1}\rightarrow g\in  SU(2)$ and Lie algebra valued  left-invariant one-forms with pull-back to $\R^{1,1}$ given by 
 \be
  \phi^*(g^{-1}\mathrm{d}g)=(g^{-1}\partial_{t}g )\,\mathrm{d}t+(g^{-1}\partial_{\sigma}g) \,\mathrm{d}\sigma
  \ee
%  with $e_i$ the Lie algebra generators of $SU(2)$, 
so to have: 
%\be\label{startingac}
%S=\textcolor{red}{\frac{1}{2}}\int_{\mathbb{R}^2}\Tr[\phi^*(g^{-1}\mathrm{d}g) \wedge \h \phi^*(g^{-1}\mathrm{d}g)],  
%\ee
\be
 S= \frac{1}{4} \int_{\mathbb{R}^2}\Tr[\phi^*(g^{-1}\mathrm{d}g) \wedge %\h 
 *\phi^*(g^{-1}\mathrm{d}g)]    \label{startingac}
\ee
where  trace is understood as the Cartan-Killing scalar product in the Lie algebra $\mathfrak{su}(2),$ and the Hodge star operator acting as $* \dd t=  \dd\sigma, *\dd\sigma= \dd t$ \footnote{We adopt the the convention $\epsilon_{01}=1$.}, yielding:
%\be
%S= \textcolor{red}{\frac{1}{2}}\int_{\mathbb{R}^2}\mathrm{d}t\mathrm{d}\sigma\ \Tr\bigl[ (g^{-1}\partial_tg)^2 - (g^{-1}\partial_{\sigma}g)^2 \bigr] \label{act0} 
%\ee
\be
S=  \frac{1}{4}\int_{\mathbb{R}^2}\mathrm{d}t\mathrm{d}\sigma\ \Tr\bigl[ \{ (g^{-1}\partial_t g)^2   - (g^{-1}\partial_{\sigma}g)^{2}  \bigr]  \label{act1}
\ee
A remarkable property of the model is that its  Euler-Lagrange equations
\begin{equation}
\partial_t(g^{-1}\partial_t g)-\partial_{\sigma}(g^{-1}\partial_{\sigma}g)=0  
\end{equation}
may be rewritten in terms of  an equivalent system of two first order partial differential equations, introducing the so called  \emph{currents}, as it is customary in the framework of integrable systems:
\begin{equation}
A^i=\Tr (g^{-1}\partial_t g) e_i, \quad {J^i}=\Tr ( g^{-1}\partial_{\sigma}g) e_i, \label{curr}
\end{equation}
%\textcolor{blue}{
%\begin{equation}
%A= (g^{-1}\partial_t g )^{i} e_i, \quad {J}=(g^{-1}\partial_{\sigma}g)^{i} e_i, \label{curr}
%\end{equation}
%}
namely, $g^{-1}\del_t g = 2 A^i e_i, g^{-1}\del_\sigma g = 2 J^i e_i$, with $\Tr \left(e_i e_j\right)=\frac{1}{2} \delta_{ij}$.  The Lagrangian  becomes:
\be
L= \frac{1}{2}\int_\R \dd\sigma (A^i\delta_{ij} A^j - J^i \delta_{ij} J^j) \label{unmodlag}
\ee
with
\begin{eqnarray} 
\partial_t A= & \partial_{\sigma} J,  \label{equiv1} \\
\partial_t J=  & \partial_{\sigma} A-[A,J].  \label{equiv2}
\end{eqnarray}
The existence of a $g \in SU(2)$ that admits the expression of the currents in the form of eq. (\ref{curr}) is guaranteed by eq.  (\ref{equiv2}), that can be read as an integrability condition. Moreover, if  the usual boundary condition for a physical field is imposed:
\be
 \lim_{\sigma\to\pm\infty} g(\sigma)=1, \label{boundco} 
 \ee
 one has that $g$ is uniquely determined from eq. (\ref{curr})

At fixed $t$, all $g$'s satisfying this boundary condition %constant at infinity  %satisfying the b.c.  $ \lim_{\sigma\to\pm\infty} g(\sigma)=1$  
form an infinite dimensional Lie group $SU(2)(\mathbb{R})\equiv \mathrm{Map}(\mathbb{R},SU(2))$,  given by smooth maps $g: \sigma\in \mathbb{R}\rightarrow g(\sigma)\in SU(2)$ which are constant at infinity. 
 Furthermore,  the currents $J$ and $A$ take values in the Lie algebra $\mathfrak{su}(2)(\mathbb{R})$ of the group $SU(2)(\mathbb{R})$ defined as the algebra of functions from $\mathbb{R}$ to $\mathfrak{g}$ that are sufficiently fast decreasing at infinity to be square-integrable.  This definition generalizes the one  of  loop algebra $\mathfrak{g}({S^1})$ and,  
 when $\mathfrak{g}$ is a semi-simple Lie algebra, the one of  Kac-Moody algebra.  

%This is a slight generalization of the definition of a loop group which is the group of smooth maps from $S^1$ to $G$. 

The carrier space of the dynamics of our system can be regarded as the tangent bundle of $SU(2)(\mathbb{R})$. Therefore, the tangent bundle description of the dynamics is given in terms of $(J,A)$ with $A$ being left generalized velocities and $J$ left configuration space coordinates. 

The infinitesimal generators of the Lie algebra $\mathfrak{su}(2)(\mathbb{R})$ can be obtained by considering the vector fields which generate the finite-dimensional Lie algebra $\mathfrak{g}$ and replacing  ordinary derivatives with functional derivatives
\begin{equation}
X_i(\sigma)=X_i^a(\sigma) \frac{\delta \ \ \ }{\delta g^a(\sigma)}.  \nonumber  % \label{funder}
\end{equation}
The following Lie bracket holds:
\begin{equation}
[X_i(\sigma),X_j(\sigma')]=c_{ij}^{\ \ k} X_k (\sigma) \delta  (\sigma-\sigma')  \nonumber 
\end{equation}
where $\sigma,\sigma' \in \mathbb{R}$.  This Lie bracket is $C^{\infty}(\mathbb{R})$-linear and 
$\mathfrak{su}(2)(\mathbb{R}) \simeq \mathfrak{su}(2) \otimes C^{\infty}(\mathbb{R})$. %is  a \emph{current algebra}.
%\item
 The real line $\mathbb{R}$ can be replaced by any smooth manifold $M$ and the Lie algebras $\mathfrak{su}(2)(M)=\mathrm{Map}(M,\mathfrak{su}(2))$ are the so called  \emph{current algebras}.

\subsection{Hamiltonian Formalism}
The target phase space is naturally given by $T^{*}SU(2)$.  Topologically it is the manifold $S^3\times \mathbb{R}^3$.
  As a group, $T^*SU(2) \simeq SU(2) \ltimes \mathbb{R}^3$, that provides a simple example of Drinfel'd double.  As a Poisson manifold, it is symplectomorphic to the group $SL(2,C)$.  
  $T^*SU(2)$ and $SL(2,\mathbb{C})$  are both Drinfel'd doubles of the group $SU(2)$ with the former named {\em classical double}.
  Let us introduce the 
conjugate variables  $(J^{i}, I_{i})$ with $J$ configuration space coordinates and $I$ left generalized momenta:
\begin{equation}
I_i=\frac{\delta L}{\delta \ (g^{-1}\partial_t g)^i}=\delta_{ij}(g^{-1}\partial_tg)^j= \delta_{ij} A^j  \nonumber \,\,\, .
\end{equation}

The Hamiltonian of the system is given by:
\begin{eqnarray}
H=\frac{1}{2}\int_{\mathbb{R}} \mathrm{d}\sigma (I_iI_j \delta^{ij} +J^i J^j \delta_{ij}) = 
\frac{1}{2}\int_{\mathbb{R}} \mathrm{d} \sigma \ I_I \ (\mathcal{H}_0^{-1})^{IJ}\ I_J   \label{undeformedH}
\end{eqnarray}
with $I_I=(I_i, J^i)$ components of the current one-form on $T^*SU(2)$ and 
\begin{equation} \label{hzero}
{(\mathcal{H}_0^{-1})}^{IJ}= \left( 
\begin{array}{cc}
\delta^{ij} & 0  \\
 0 & \delta_{ij} 
\end{array} 
\right)
\end{equation} 
is a Riemannian metric on $T^*SU(2).$

The  Hamiltonian description of the Principal Chiral Model on $SU(2)$ naturally involves the Riemannian {\em generalized} metric $\mathcal{H}_0^{-1}$  on the cotangent bundle \cite{Witten1983, RB1984}.

The first-order Lagrangian, together with the canonical one-form and the symplectic form, allows to determine the
equal-time Poisson brackets \cite{Witten1983, RB1984}:
%and the equations of motion \ref{equiv1} and \ref{equiv2} can be recovered considering a particular form for the Poisson brackets:
\begin{eqnarray} 
\{I_i(\sigma),I_j(\sigma')\} &= &{ \epsilon_{ij\;}}^k I_k(\sigma)\delta(\sigma-\sigma') \nonumber \\
\{I_i(\sigma),J^j(\sigma')\} & = & {\epsilon_{ki\;}}^j J^k(\sigma) \delta(\sigma-\sigma')-\delta_{i}^j\delta'(\sigma-\sigma') \label{Poisson} \\
\{J^i(\sigma),J^j(\sigma')\} & = & 0  \nonumber 
\end{eqnarray}
The  $I$'s  are generators of the affine Lie algebra $\mathfrak{su}(2)(\mathbb{R})$, while the $J$'s span an Abelian algebra $\mathfrak{a}(\mathbb{R})$.
$I$ and $J$ span the infinite-dimensional  current algebra $\mathfrak{c}_1=\mathfrak{su}(2)(\mathbb{R})\ltimes \mathfrak{a}(\mathbb{R})$.  Their corresponding equations of motion
 are given by:
\begin{eqnarray}
 \partial_t I_j(\sigma) & = &  \{H,I_j (\sigma)\}=
%\frac{1}{2} \int_{\mathbb{R}} \mathrm{d}\sigma \ \bigl(\{I_i(\sigma),I_j(\sigma')\} I_k(\sigma)+I_i(\sigma)\{I_k(\sigma),I_j(\sigma')\}+ \\
%& \qquad \qquad+ \left(\{J^p(\sigma),I_j(\sigma')\} J^q(\sigma)+J^p(\sigma)\{J^q(\sigma),I_j(\sigma')\} \right)\delta_{pi}\delta_{qk}\bigr)\delta^{ik}=\\ 
%&=  \int_{\mathbb{R}}\mathrm{d}\sigma \ \bigl[{\epsilon_{ij}}^k I_k(\sigma)\delta(\sigma-\sigma') I_i(\sigma) +  \bigl(\epsilon_{ij k} J^k(\sigma) \delta(\sigma-\sigma')-\delta_{ij}\delta'(\sigma-\sigma')\bigr)J^i(\sigma) \bigr]= \\
%&=
  \partial_{\sigma}J^k \delta_{kj}(\sigma)   \\  \label{eomI}
%\end{split}
\partial_t J^j(\sigma) & = & \{H,J^j(\sigma)\}=\partial_{\sigma}I_k \delta^{kj}(\sigma) -{\epsilon^{\ jl}}_{k}I_lJ^k(\sigma).    \label{eomJ}
\end{eqnarray}
%\end{itemize}
%}

%\frame{
%\begin{itemize}

%\pause
%\item The carrier space of the dynamics is the cotangent bundle of $G(\mathbb{R})$, with   the currents $(J^i, I_i)$ playing the role of conjugate variables with $I$ the left generalized momenta and $J$ the  left configuration space coordinates. 

%\end{itemize}
%}

%\frame{
%\begin{itemize}
It is possible to give an equivalent description of the dynamics in terms of a new one-parameter family of Poisson algebras and modified Hamiltonians, with the currents playing a symmetric role \cite{R89, R892}. 
In particular, it is possible to
deform the Poisson brackets in eq. (\ref{Poisson}) by introducing a purely imaginary parameter $\tau$ as follows:
\begin{eqnarray} 
\!\!\!\!\!\!\!\!\!\!\!\!\{I_i(\sigma),I_j(\sigma')\} & = & (1-\tau^2){\epsilon_{ij}}^k I_k(\sigma)\delta(\sigma-\sigma') \nonumber  \\
\!\!\!\!\!\!\!\!\!\!\!\!\{I_i(\sigma),J^j(\sigma')\} & =& (1-\tau^2)\bigl(J^k(\sigma){ \epsilon_{ki}}^j  \delta(\sigma-\sigma')-(1-\tau^2)^2\delta_{i}^j\delta'(\sigma-\sigma')) \nonumber \\
\!\!\!\!\!\!\!\!\!\!\!\!\{J^i(\sigma),J^j(\sigma')\} & = & (1-\tau^2)\tau^2 {\epsilon^{ij}}_k I_k(\sigma)\delta(\sigma-\sigma').  \nonumber
\end{eqnarray}
The factor ($1-\tau^2)$ is never zero for imaginary $\tau$.

The new brackets correspond to the infinite-dimensional Lie  algebra $\mathfrak{c}_2\simeq\mathfrak{sl}(2,\mathbb{C})(\mathbb{R})$ %which%, for imaginary $\tau$, %our choice from now on,  can be easily recognized to be 
%is 
isomorphic to the current algebra modelled on the Lorentz algebra $\mathfrak{sl}(2,\mathbb{C})$. %hat is  $\mathfrak{c}_2\simeq\mathfrak{sl}(2,\mathbb{C})(\R)$\footnote{For real $\tau$ it is instead isomorphic to the algebra $\mathfrak{so}(4)(\R)$.  
%The latter   case is the one analyzed in detail in  \cite{R89, RSV93, RSV96} with respect to quantization and integrability. Here we stick  to imaginary $\tau$, this being the choice which unveils  the double group structure. }.  
%The Lie algebra $\mathfrak{c}_1=\mathfrak{su}(2)(\mathbb{R})\ltimes \mathfrak{a}$  recovered when $\tau \rightarrow 0$.

The modified Hamiltonian is correspondingly:
\begin{equation}
H_{\tau}=\frac{1}{2(1-\tau^2)^2}\int_{\mathbb{R}} \mathrm{d}\sigma \  (I_iI_j \delta^{ij}+J^i J^j \delta_{ij}) \label{modiha} \nonumber
\end{equation}
%and, in the limit $\tau \rightarrow 0$, the algebra and the Hamiltonian reduce to the original  ones.  
while the  equations of motion remain unmodified.
%\begin{eqnarray}
 %\partial_t I_j(\sigma)&=&  \{H_\tau,I_j(\sigma)\}= \partial_{\sigma}J^k \delta_{kj}  \nonumber \\ 
%\partial_t J^j(\sigma)&=& \{H_\tau,J^j(\sigma)\}=\partial_{\sigma}I_k \delta^{kj}  - %(1-\tau^2) 
%{\epsilon^{\ jl}}_{k}I_lJ^k.  \nonumber
%\end{eqnarray}
Hence, one gets an alternative description of one and the same dynamics even considering deformed algebras of $SL(2, \mathbb{C})$.

The new equations of motion read then as:
\beqa
 \partial_t I_j(\sigma)&=&  \{H_\tau,I_j(\sigma)\}= \partial_{\sigma}J^k \delta_{kj}\label{newone}\\ 
\partial_t J^j(\sigma)&=& \{H_\tau,J^j(\sigma)\}=\partial_{\sigma}I_k \delta^{kj}  - {\epsilon^{\ jl}}_{k}I_lJ^k. \label{newtwo}
\eeqa
which coincide with   eqs. \eqn{eomI}, \eqn{eomJ}. 

 The fields $I_{i}$ and $J^{i}$ can be rescaled according to: 
\begin{equation}
\frac{I_i}{1-\tau^2}\rightarrow I_i\;\;\; ; \;\;\;\frac{J^i}{1-\tau^2}\rightarrow J^i.  \nonumber %\label{redef}
\end{equation} 
%Since now on, we shall keep the same notation for the rescaled fields unless otherwise stated. 
The rescaled Hamiltonian  $H_{\tau}$ becomes identical to the undeformed one $H$, while the Poisson algebra acquires the form:
\begin{eqnarray}
\{I_i(\sigma),I_j(\sigma')\}&= &{\epsilon_{ij}}^k I_k(\sigma)\delta(\sigma-\sigma'),   \nonumber %\label{modipoi1} 
\\
\{I_i(\sigma),J^j(\sigma')\}&= &J^k(\sigma){ \epsilon_{ki}}^j  \delta(\sigma-\sigma')-\delta_{i}^j\delta'(\sigma-\sigma'),  \nonumber %\label{modipoi2}
\\
\{J^i(\sigma),J^j(\sigma')\}& = & \tau^2 {\epsilon^{ijk}} I_k(\sigma)\delta(\sigma-\sigma'). \nonumber \label{modipoi3}
\end{eqnarray}

 New generators  can be introduced at this point with the aim of showing the bi-algebra structure of $\mathfrak{sl}(2, \mathbb{C})(\mathbb{R})$. %structure of the Lie algebra described by the deformed Poisson brackets.  
 Keeping the  generators of  $\mathfrak{su}(2)(\mathbb{R})$ unmodified, one can consider the  linear combination:
 \begin{equation}
 K^i(\sigma)=J^i(\sigma)-i\tau {\epsilon^{\ell i3}}I_{\ell}(\sigma).  \nonumber 
\end{equation}
From  the deformed Poisson brackets % \eqn{modipoi1}-\eqn{modipoi3}
 it is possible to derive the ones of the new generators:
%\begin{equation}
%\begin{split}
%& \{K_i(\sigma),K_j(\sigma')\}= \{J_i(\sigma)-\epsilon_{ik3}I_k(\sigma),J_j(\sigma')-\epsilon_{jl3}I_l(\sigma')\}= \\
%&=  \{J_i(\sigma),J_j(\sigma')\}-i\tau\epsilon_{jl3}\{J_i(\sigma),I_l(\sigma')\}-i\tau \epsilon_{ik3}\{I_k(\sigma),J_j(\sigma')\}+ \\
%& \quad +\tau^2 \epsilon_{ik3}\epsilon_{jl3}\{I_k(\sigma),I_l(\sigma')\}= \\
%& = (1-\tau^2)\tau^3\delta(\sigma-\sigma')[(-\epsilon_{ijk}+\epsilon_{is3}\epsilon_{jl3}\epsilon_{slk})\tau I_k(\sigma')-i(\epsilon_{il3}\epsilon_{ljk}-\epsilon_{jl3}\epsilon_{lik})J_k(\sigma')],  
%\end{split}
%\end{equation}
%so we can write 
\begin{equation}
%\begin{split}
\{K^i(\sigma),K^j(\sigma')\}  
%&  (1-\tau^2)\tau^3\delta(\sigma-\sigma')[-\tau \epsilon_{ijl}\epsilon_{lk3}\epsilon_{ks3}I_s(\sigma')+i\epsilon_{ijl}\epsilon_{lk3}J_k(\sigma')]= \\
= i \tau {f^{ij}}_{k} K^k(\sigma') \,  \delta(\sigma-\sigma')   \nonumber %\label{poisk}
%\end{split}
\end{equation}
showing that $K$'s span the $\mathfrak{sb}(2,\mathbb{C})(\mathbb{R})$ Lie algebra, with structure constants ${f^{ij}}_k=\epsilon^{ij\ell}\epsilon_{ \ell 3k}$, while for the mixed Poisson brackets one finds: 
\begin{equation}
 \{I_i(\sigma),K^j(\sigma')\} \!\! = \!\!  \left(K^k(\sigma'){\epsilon_{k i}}^{\,j}-i \tau   I_k(\sigma') {f^{kj}}_{i}  \,\,\,%\epsilon^{kjs}{\epsilon_{s3i}}
\right) 
\delta(\sigma-\sigma')
- \delta_{i}^j\delta'(\sigma-\sigma') .  \nonumber
 \end{equation}
We refer to $\mathfrak{sb}(2,\mathbb{C})$ as the Lie algebra of $SB(2,\mathbb{C})$, the Borelian subgroup of $SL(2,\mathbb{C})$ of complex $2 \times 2$ upper triangular matrices with unit determinant and real diagonal.
In this way, the Lie algebra $\mathfrak{c}_2\equiv\mathfrak{sl}(2,\mathbb{C})(\mathbb{R})$ has been expressed as a Drinfel'd double algebra $\mathfrak{c}_2 =\mathfrak{su}(2)(\mathbb{R}) 
\Join\mathfrak{sb}(2,\mathbb{C})(\mathbb{R})$, in the sense of a Manin triple decomposition.
%\item
%$SL(2,\mathbb{C})$ is the Drinfel'd double group of $SU(2)$ and $SB(2, \mathbb{C})$.
% \item
 %\textcolor{red}{questa e' un'ipotesi, andrebbe provato}. 

%up to a central extension with central charge equal to $-1$, i.e. just like the affine algebra associated with the Drinfel'd double of the Lie algebra $\mathfrak{su}(2)$ considered at the beginning.
 %\{I_i(\sigma),J^j(\sigma')-i\tau \epsilon^{jl3}I_l(\sigma')\} \nonumber\\
 %  \nonumber
%\label{poiski}
%\end{eqnarray} 
%where the structure constants of the Lie algebra $\mathfrak{sb}(2,\C)$, $ \epsilon^{kjs}{\epsilon_{s3i}}={f^{kj}}_i$, are recognized. 
On rewriting the alternative Hamiltonian in eq. (\ref{modiha}) in terms of the new generators, the $SU(2)$ chiral model is completely described  by the one-parameter family of Hamiltonian functions 
\begin{equation}
H_{\tau}=\frac{1}{2}\int_{\mathbb{R}} \mathrm{d}\sigma \  \left[I_s I_{\ell}\left( \delta_i^s\delta_j^{\ell} -\tau^2 {\epsilon^s}_{i3} {\epsilon^{\ell}}_{j3}\right)\delta^{ij}+ K^i K^j \delta_{ij} + 2i\tau {\epsilon}^{s \ell 3} I_s K^q \delta_{\ell q}\right] \label{modiha2}
\end{equation}
with Poisson brackets given by:
\begin{eqnarray}
\{I_i(\sigma),I_j(\sigma')\}&= & {\epsilon_{ij}}^k I_k(\sigma)\delta(\sigma-\sigma') \label{IIcur} \\
\{K^i(\sigma),K^j(\sigma')\}&=&  i \tau{f^{ij}}_kK^k(\sigma') \delta(\sigma-\sigma')\label{KKcur}\\
\{I_i(\sigma),K^j(\sigma')\} &=&\left(K^k(\sigma'){\epsilon_{ki}}^j+i \tau   {f^{jk}}_i I_k(\sigma') \right) \delta(\sigma-\sigma')-\delta_{i}^j\delta'(\sigma-\sigma') \label{IKcur}
\end{eqnarray}
yielding the interesting  result that the Principal Chiral Model with compact target space may be described in terms of a non-compact current algebra. 
This result can be traced back to   the fact that the cotangent bundle of the group $SU(2)$ is symplectomorphic to the group $SL(2,\mathbb{C}).$ 

\section{Drinfel'd Doubles and Manin triples }

As a group $T^*SU(2)$ is the semi-direct product $SU(2) \ltimes \mathbb{R}^3$  with Lie algebra $\mathfrak{su}(2) \ltimes \mathbb{R}^3$ and Lie brackets given by:
%$\;$\\
$$
\left[L_i,L_j\right]  =  {\epsilon_{ij}}^k L_k \label{JJ} ~~~~~~
\left[T_i,T_j\right] =  0 \label{PP}~~~~~~
\left[L_i,T_j\right]= {\epsilon_{ij}}^k T_k  \,\,\, .
$$
Here $L_i, T_i,$ $ i = 1, 2, 3$ generate respectively the algebra $\mathfrak{su}(2)$ and $ \mathbb{R}^3$.
% The  non-trivial Poisson brackets in eq.s (\ref{}) are the Kirillov-Souriau-Konstant (KSK) brackets on  the dual algebra $\mathfrak{\tilde{g}}$ .

According to ref.s \cite{R89, MSS92, RSV93} the carrier space of the dynamics  can be generalized to   $SL(2, \mathbb{C})$, the Drinfel'd double of $SU(2)$, which, roughly speaking, can be obtained by deforming  the Abelian subgroup $\mathbb{R}^3$ of the semi-direct product above. % and  a similar generalization has been proposed  for the Principal Chiral Model  \cite{R89} and  the Wess-Zumino-Witten Model \cite{RSV93}.

Let us give the definition of $SL(2,\mathbb{C})$ as a Drinfel'd double by starting from the properties of its Lie algebra.

The Lie algebra $\mathfrak{sl}(2,\mathbb{C)}$ %can be regarded as a real form of the complex Lie algebra $\mathfrak{sl}(2)$ %Indeed, $\mathfrak{sl}(2)$ 
 is defined by the Lie brackets:
 \begin{equation}
%\begin{aligned}
%{} &
 [e_i, e_j]=i {\epsilon_{ij}}^k e_k  \,\,\,\,;\,\,\,\, %\\ &
[e_i, b_j]=i {\epsilon_{ij}}^k b_k  \,\,\,\,;\,\,\,\, %\\ &
[b_i, b_j]=-i {\epsilon_{ij}}^k e_k.   \nonumber
%\end{aligned}
\end{equation}
%\textcolor{blue}{
%\begin{equation}
%\begin{aligned}
%{} & 
%[t_1, t_2]=t_3 %\\ & 
%\,\,\,\,\, 
%[t_2, t_3]=2t_2 % \\ &
%\,\,\,\,\,\,
%[t_3, t_1]=2 t_1    \nonumber
%\end{aligned}
%\end{equation}}
%with : %$t_{i}$ %its generators:
%\begin{equation}
%t_1=\begin{pmatrix}
%  0 & 1 \\
%  0 & 0
% \end{pmatrix}; \quad t_2=\begin{pmatrix}   
%  0 & 0 \\
%  1 & 0
% \end{pmatrix} ; \quad t_3=\begin{pmatrix}   
%  1 & 0 \\
%  0 & -1
 %\end{pmatrix}   \nonumber
%\end{equation}
%\pause
%\item
% By taking complex linear combinations %of the $\mathfrak{sl}(2)$ generators given by
with
\begin{equation} 
e_i=\frac{\sigma_i}{2} %, \quad e_2=\frac{\sigma_2}{2}, \quad e_3=\frac{\sigma_3}{2}  
\,\,\,\,\, \mbox{
 generators of $\mathfrak{su}(2)$} \nonumber
\end{equation}
\begin{equation}
b_i=i e_i \,\,\,\,\,\,\,\,\,\,\quad i=1,2,3   \nonumber
\end{equation}
%\begin{equation}
%\label{liesl2c}

It is equipped with two non-degenerate invariant scalar products: 
\begin{equation}
\langle u,v \rangle =2{\rm Im}(\mbox{Tr}(uv)) \quad \forall u,v \in \mathfrak{sl}(2,\mathbb{C})  \label{scapro1}
\end{equation}
\begin{equation}
(u,v)=2{\rm Re}(\mbox{Tr}(uv)) \quad \forall u,v \in \mathfrak{sl}(2,\mathbb{C}). \label{scapro2}
\end{equation}
The scalar product $ \langle u,v \rangle$  defines two maximally isotropic subalgebras of $\mathfrak{sl}(2,\mathbb{C})$ \begin{equation}
[e_i, e_j]= i {\epsilon_{ij}}^k e_k, ~~~~~[\tilde e^i,  e_j]= i{\epsilon_{jk}}^i \tilde e^k+ i e_k {f^{ki}}_j , ~~~~~[\tilde e^i, \tilde e^j]= i {f^{ij}}_k \tilde e^k \label{algebra}
\end{equation}
spanned by $\{e_i\}$ and by the linear combination
 % \label{etilde}
$\tilde e^i= b_i -\epsilon_{ij3}e_j  \nonumber
$
corresponding, respectively, to 
%It is important to note that 
%\textcolor{red}{
 $\mathfrak{su}(2)$ and $\mathfrak{sb}(2,\mathbb{C})$ which are maximally isotropic with respect to  the scalar product $\langle u,v \rangle$ since:
\begin{equation}
\langle e_i, e_j \rangle=0, \quad \langle \tilde{e}^i, \tilde{e}^j \rangle=0, \quad \langle e_i, \tilde{e}^j \rangle = \delta^{j}_{i}\nonumber
\end{equation}

Each subalgebra acts on the other one non-trivially by co-adjoint action according to the algebra in (\ref{algebra}).

Furthermore, the Lie bracket
\begin{equation}
%\label{liesusb}
[\tilde{e}^i, e_j]=i {\epsilon^i}_{jk} \tilde{e}^k+i {f^{ki}}_j e_k  \nonumber
\end{equation}
puts in evidence that
%}
$SL(2,\mathbb{C})$ is the Drinfel'd double of $SU(2)$ and $SB(2, \mathbb{C})$ with polarization $\mathfrak{sl}(2,\mathbb{C})=\mathfrak{su}(2) \bowtie \mathfrak{sb}(2,\mathbb{C})$ and therefore $(\mathfrak{sl}(2,\mathbb{C}), \mathfrak{su}(2), \mathfrak{sb}(2,\mathbb{C}))$ is a Manin triple. Hence, $SU(2)$ and $SB(2,\mathbb{C})$ are dual groups. %Observe that the Lie brackets of (\ref{liesl2c}) (for the $e_i$), (\ref{liesb}) and (\ref{liesusb}) have exactly the form as (\ref{drinfbrack}).
The following 
 {\it doubled} notation can be introduced:
\begin{equation}
e_I= \left( 
\begin{array}{c}
e_i\\ {\tilde e}^i 
\end{array} \right), 
 \qquad  e_i \in \mathfrak{su}(2), \quad  {\tilde e}^i \in \mathfrak{sb}(2,\mathbb{C}) \,\, .  \nonumber
\end{equation}
The scalar product (\ref{scapro1}) then gives:
 \begin{equation}
(e_I,e_J)=\eta_{IJ}=  \left(
\begin{array}{cc} 
0 & \delta_i^j \\ \nonumber
\delta_j^i &  0
\end{array}  \right) 
\end{equation}
which is  $O(3,3)$ invariant  by construction. 

Instead, the scalar product (\ref{scapro2}) yields:
\begin{equation}
(e_I,e_J)=  \left(
\begin{array}{cc}
\delta_{ij} & \epsilon_{i p 3}\delta^{pj} \\  \nonumber 
 \delta^{ip}\epsilon_{j p 3}&  \delta^{ij}- \epsilon^{i k3} \delta_{kl}  \epsilon^{j  l3} \,\,\,   \nonumber
\end{array} \right) \,\,\, .
\end{equation}

The splitting ${\mathfrak d}= C_{+} \oplus C_{-}$ with $C_+, C_-$ spanned by $\{e_i\}$, $\{b_i\}$ respectively,  defines a positive definite metric $\mathcal{H}$ on ${\mathfrak d}$ via:
\begin{equation}
\mathcal{H}= (\;,\;)_{C_+}-  (\;,\;)_{C_-}  \,\,\,  \label{2.30}
\end{equation}
%It is immediate to check that the metric ${\mathcal H}$,  that will be indicated since now on by  double round brackets: 
%$$
%((e_i,e_j)):= (e_i,e_j); ~~~~~ ((b_i,b_j)):=-(b_i,b_j);~~~~~ ((e_i,b_j)):= (e_i,b_j)=0 
%$$
satisfying
$$ {\mathcal H}^T \eta {\mathcal H} = \eta \,\,\,  $$ 
namely ${\mathcal H}$  is a pseudo-orthogonal $O(3,3)$ metric and $\eta$ is the $O(3,3)$ invariant metric. 

It is worth to observe here that in this framework the two structures ${\mathcal H}$ and $\eta$, that play the same role as the generalized metric and the $O(d,d)$ invariant metric in the T-dual invariant formulation of string theory, naturally emerge out.

%\item
%There is also another \textcolor{blue}{non-degenerate invariant scalar product} which can be defined on \textcolor{blue}{$\mathfrak{sl}(2,\mathbb{C})$} as:
%\textcolor{red}{
%\begin{equation}
%\label{sp2}
%(v,w)=2 \text{Re}\left[\text{Tr}\left(vw \right) \right], \quad \forall \, v,w \in \mathfrak{sl}(2,\mathbb{C}).  \nonumber
%\end{equation} }
%$\mathfrak{su}(2)$ and $\mathfrak{sb}(2,\mathbb{C})$ are no longer isotropic subspaces with respect to this scalar product, in fact, for the basis elements:
%\begin{equation}
%(e_i, e_j)=\delta_{ij}, \quad (b_i, b_j)=-\delta_{ij}, \quad (e_i, b_j)=0.  \nonumber
%\end{equation}
%not giving rise to a positive-definite metric.  
%\pause
%\item
%On $C_+$, $C_-$ respectively the two subspaces spanned by $\{e_i \}$ and $\{ b_i\}$, the splitting \textcolor{blue}{$\mathfrak{sl}(2,\mathbb{C})=C_+ \oplus C_-$ } defines a positive definite metric  \textcolor{red}{$\mathcal{H}$ on $\mathfrak{sl}(2,\mathbb{C})$}:
%\begin{equation}
%\mathcal{H}=\left( , \right)_{C_+}-\left( , \right)_{C_-}.  \nonumber
%\end{equation}
%This is a \textcolor{blue}{ Riemannian metric  $\left( \left(	\, ,\,  \right) \right)$ }: %In particular:
%\textcolor{red}{
%\begin{eqnarray}
%\label{riemannianmetric}
%\left( \left(e_i, e_j \right) \right) \equiv  \left(e_i, e_j \right), \quad \left( \left( b_i, b_j \right) \right) \equiv -\left(b_i, b_j \right) \nonumber \\
 %\quad \left( \left(e_i, b_j \right) \right) \equiv \left(e_i ,b_j\right)=0.  \nonumber
%\end{eqnarray}}
%\end{itemize}
%}

\section{Family of models $\tau$-dependent}
Rewriting $H_{\tau}$ in eq. (\ref{modiha2}) in terms of $I_{J} = (I_{j}, K^{j})$ shows that 
there exists  a whole  family of models, described by the Hamiltonians labelled by the parameter $\tau$
\begin{equation}
H_{\tau}=\frac{1}{2}\int_{\mathbb{R}} \mathrm{d}\sigma \, { I}_L ({{\mathcal H}^{-1}_\tau})^{LM} { I}_M  \nonumber
\end{equation}
with  ${\mathcal H}_{\tau}^{-1}$ being the {\em Riemannian generalized metric} %which we choose to denote as an  inverse metric, 
% ${\mathcal H}_{\tau}^{-1}$:
\begin{equation} % \label{Htau-1}
{\mathcal H}_{\tau}^{-1}=  \left(
\begin{array}{cc}
 h^{ij}(\tau) &i\tau \epsilon^{ip3}\delta_{pj} \\  
  i\tau \delta_{ip}\epsilon^{jp3} &  \delta_{ij}  \label{Htauu}
\end{array}  \right) 
\end{equation}
where: % it has been defined, for  future convenience:
\begin{equation}
\label{sb2cmetric}
h^{ij}(\tau)= \delta^{ij}-\tau^2 \epsilon^{ia3}\delta_{ab} \epsilon^{jb3} \,\,\, . \nonumber % \label{htau}
\end{equation}
They  are related (and indeed equivalent) to the standard $SU(2)$ chiral model  by the $O(3,3)$ transformation $K^i(\sigma)=J^i(\sigma)-i\tau {\epsilon^{li3}}I_l(\sigma)$, symmetry of the dynamics because it maps solutions into solutions. 
%\item
%It puts in evidence the bialgebra structure of  $\mathfrak{sb}(2,\mathbb{C})$.

It is worth to observe how naturally the two structures $\eta$ and $\mathcal{H}$ naturally appear in this context just as they naturally appear in a T-dual invariant formulation of the string world-sheet.

\section{Born Geometry and $B$-transformations}

Eq. (\ref{undeformedH})  shows that the original undeformed Hamiltonian $H$ naturally involves ${(\mathcal{H}_0^{-1})}^{IJ}$,
%\begin{equation} \nonumber %\label{compactH}
%H=\frac{1}{2}\int_{\mathbb{R}} \mathrm{d} \sigma \ I_I \ (\mathcal{H}_0^{-1})^{IJ}\ I_J,  
%\end{equation}
%where $I_I=(I_i, J^i)$ are components of the current 1-form on $T^*SU(2)$ and 
%\begin{equation} \nonumber % \label{hzero}
%{(\mathcal{H}_0^{-1})}^{IJ}=
%\left(
%\begin{array}{cc}
%\delta^{ij} & 0 \\
%0 & \delta_{ij}
%\end{array}
%\right)     \label{H0}
%\end{equation}
a Riemannian metric on $T^{*}SU(2)$. Such metric  can be interpreted as one of the structures defining a left-invariant Born geometry on $T^{*}SU(2)$. In this case the transformation defining a Born geometry acts as an O(3,3) transformation of the target phase $T^{*}SU(2)$.

%\item
% The transformation defining a Born geometry acts as an $O(3,3)$  transformation of  the target phase  $T^*SU(2)$.
 $T^*SU(2)$ is a Drinfel'd double with Lie algebra $\mathfrak{su}(2) \ltimes \mathbb{R}^3.$  Such Lie algebra has a natural (symmetric, non-degenerate) pairing $<{\cdot \ ,\ \cdot }>$ such that $\mathfrak{su}(2)$ and $\mathbb{R}^3$ are maximally isotropic subspaces with respect to it. Moreover, $\mathfrak{su}(2) \ltimes \mathbb{R}^3$ can be seen as a split vector space $\mathfrak{su}(2) \oplus \mathbb{R}^3,$ thus it can be naturally endowed with a para-complex structure $\kappa,$ i.e. $\kappa \in \mathrm{End}(\mathfrak{su}(2) \ltimes \mathbb{R}^3)$ such that $\kappa^2 =1$ with $\mathfrak{su}(2)$ eigenspace of $\kappa$ associated with the eigenvalue $+1$ and $\mathbb{R}^3$ eigenspace associated with the eigenvalue $-1.$ The structures  $<{\cdot \ ,\ \cdot }>$ and $\kappa$ satisfy a compatibility condition %$$<{\kappa(\xi), \psi}>=-<\kappa(\psi), \xi>, \hspace{1cm} \forall \xi, \psi \in \mathfrak{su}(2)\ltimes \mathbb{R}^3,$$ 
 which defines a two-form $\omega$ on $\mathfrak{su}(2) \ltimes \mathbb{R}^3.$ %Summarizing, $(<\cdot \ ,\ \cdot > \ \kappa)$ related by the above compatibility condition define a para-Hermitian structure on  $\mathfrak{su}(2) \ltimes \mathbb{R}^3.$

 %$T^{*}SU(2)$ is a {\em Born manifold}, i.e. a phase space equipped with a para-complex structure ($\eta,  \kappa, \mathcal{H}_0$) with $\kappa \in \mathrm{End}(\mathfrak{su}(2) \ltimes \mathbb{R}^3)$ such that $\kappa^2 =\mathbb{1}$ with $\mathfrak{su}(2)$ eigenspace of $\kappa$ associated with the eigenvalue $+1$ and $\mathbb{R}^3$ eigenspace with the eigenvalue $-1.$ The structures  $<\cdot \ ,\ \cdot >$ and $\kappa$ satisfy a compatibility condition $$<\kappa(\xi), \psi>=-<\kappa(\psi), \xi>, \hspace{1cm} \forall \xi, \psi \in \mathfrak{su}(2)\ltimes \mathbb{R}^3,$$ which defines a two-form $\omega$ on $\mathfrak{su}(2) \ltimes \mathbb{R}^3.$ %Summarizing, $(< \cdot \>, k)$  % \braket{\cdot \ ,\ \cdot }, \ \kappa)$
% related by the above compatibility condition define a para-Hermitian structure on  $\mathfrak{su}(2) \ltimes \R^3.$

The defining relations for the Born structure $(\eta , \kappa, \mathcal{H}_{0})$ on $T^{*}SU(2)$ are given by: 
$$\eta^{-1} \mathcal{H}_0=  \mathcal{H}_0^{-1} \eta \hspace{2cm} \omega^{-1} \mathcal{H}_0= - \mathcal{H}_0^{-1} \omega.$$

This is the canonical Born geometry induced by the Drinfel'd double structure \cite{Borngeomf, marottaszabo}.

The deformed Hamiltonian $H_\tau$ in eq. (\ref{Htauu}) also gives a Riemannian metric on $T^*SU(2)$ and is a $B$-transformation of the metric $\mathcal{H}_0.$  

Let us introduce the   $\tau$-dependent $B$-transformation 
\begin{equation} % \label{onepara}
e^{B(\tau)}= \left(
\begin{array}{cc}
1 & i\tau B \\  \nonumber 
0 & 1
\end{array}  \right) 
\in O(3,3) 
\end{equation}
with components of the tensor $B$ given by $B^{ij}= \epsilon^{ij3}$
% \footnote{One may regard it as $B= \epsilon^{ij3}X_j \otimes \tilde{\theta}_i.$} and $\tau \in \mathbb{C}.$ Note that, for $i\tau$ real, the $B$-transformations in the form \eqn{onepara} close a \textcolor{blue}{one-parameter} subgroup of $O(3,3),$ with the identity given by $e^B(0) = \mathds{1}$ and the inverse obtained replacing $i\tau$ with $-i\tau.$

 $\mathcal{H}_\tau$  is then obtained by the  $B$-transformation acting on  $\mathcal{H}_0:$  
\begin{equation}  \label{Htrans}
\mathcal{H}_\tau= \bigl(e^{-B(\tau)}\bigr)^t \mathcal{H}_0 e^{B(\tau)}. \nonumber 
\end{equation}
i.e. it has components:
\begin{equation}
{({\mathcal H}_\tau})_{IJ}=
%\begin{matrix}
\left( \begin{array}{cc} 
 \delta_{ij}  & i\tau \delta_{ip}\epsilon^{jp3}  \\
i\tau \epsilon^{ip3}\delta_{pj} & \delta^{ij}- \tau^2\epsilon^{is3} \delta_{sl} \epsilon^{j l3.}
\end{array}
\right)
%\end{matrix}
\end{equation}

Furthermore, the left-invariant para-Hermitian structure $(\eta, \kappa)$ is transformed under $e^{B(\tau)}.$ In particular, the only structure which changes under such transformation is the para-complex structure $\kappa,$ i.e. 
\begin{equation}
\kappa_{\tau}= e^{B(\tau)} \kappa e^{-B(\tau)}
\end{equation}
with  $\kappa_\tau$  still compatible with $\eta$,
%\frame{
%\begin{itemize}
%\item
%The left-invariant para-Hermitian structure $(\eta, \kappa)$ is transformed under $e^{B(\tau)}.$  The only structure which changes is the para-complex structure $\kappa,$ i.e. 
%\begin{equation}
%\kappa_{\tau}= e^{B(\tau)} \kappa e^{-B(\tau)}  \nonumber
%\end{equation}  
%with  $\kappa_\tau$  still compatible with $\eta,$
 so that the fundamental two-form becomes $\omega_{\tau}= \eta \kappa_\tau.$ 
%\item
In matrix form, the new almost para-Hermitian structure reads as:
\begin{equation}
\kappa_{\tau}=
\left( 
\begin{array}{cc}
1 & 2 i\tau B \\
0 & 1                 
\end{array}        \nonumber
\right)          
\hspace{1cm} 
\eta= 
\left(
 \begin{array}{cc}
0 & 1 \\
1 & 0
\end{array}
\right)
\hspace{1cm}
 \omega_\tau=
\left(
\begin{array}{cc}
0 & 1 \\
1 & 2i \tau \eta B
\end{array}
\right)
\end{equation}
where $\eta B \in \Gamma(\wedge^2 T^* \mathbb{R}^3).$ %Note that the new almost para-Hermitian structure still has $TSU(2)$ as eigenbundle while $T\R^3$ is transformed in a non-involutive distribution $V_{\tau}$ whose sections are generated by vector fields in the form ${\bar Y}^i= Y^i + i \tau \epsilon^{ij3}X_j.$
%We can easily check that 
%\item
%The metric $\mathcal{H}_\tau$ gives a Born structure on $T^*SU(2)$ together with $(\eta, \kappa_\tau),$ for each value of the parameter $\tau.$ 
%In this sense we obtain a family of Born structures from the deformation of the current algebra.

The family of equivalent Hamiltonian descriptions of the $SU(2)$ PCM  can be understood in terms of a one-parameter family of Born geometries for $T^*SU(2)$,  corresponding, for each choice of the parameter $\tau$,  to a specific splitting of the phase space, with the value $\tau=0$  the canonical splitting.

\section{Poisson-Lie symmetry: Definition}

A coordinate  transformation associated to a Lie group ${\cal G}$ is a {\em symmetry} of the dynamics  if it leaves the equations of motion unchanged in form. 
If it is also a symmetry for the auxiliary geometric structures  which are relevant to the chosen formulation
(symplectic form,  Poisson brackets and Hamiltonian in the Hamiltonian approach, functional action and Lagrangian in the Lagrangian approach), namely an {\em isometry}, then the symmetry yields constants of motion, which are associated to conserved N\"other currents. 

If a given symmetry of the dynamics under the Lie group ${\cal G}$ is not a symmetry for the auxiliary geometric structures, that is not an isometry,  but the violation is governed by the Maurer-Cartan structure equation of the Drinfel'd double %${\cal \tilde{G}}$ 
associated to  ${\cal G}$, the symmetry of the dynamics is a  Poisson-Lie symmetry \cite{BMPV}.

%}

%\frame{
%\begin{itemize}

The PCM described by
\begin{equation}
H_{\tau}=\frac{1}{2}\int_{\mathbb{R}} \mathrm{d}\sigma \  \left[I_sI_l ({{\mathcal H}_\tau}^{-1})^{sl} + K^s K^l ({{\mathcal H}_\tau}^{-1})_{sl} + K^s I_l  {{( {{\mathcal H}_\tau}^{-1} )}_s}^{l} + I_s  K^l{ ({{\mathcal H}_\tau}^{-1})^{s}}_l \right]  \,\,\, ,  \nonumber  \label{modiha3}
\end{equation}
 together with the Poisson algebra
  \begin{eqnarray}
\{I_i(\sigma),I_j(\sigma')\}&= & {\epsilon_{ij}}^k I_k(\sigma)\delta(\sigma-\sigma') \label{IIcur} \\
\{K^i(\sigma),K^j(\sigma')\}&=&  i \tau{f^{ij}}_kK^k(\sigma') \delta(\sigma-\sigma')\label{KKcur}\\
\{I_i(\sigma),K^j(\sigma')\} &=&\left(K^k(\sigma'){\epsilon_{ki}}^j+i \tau   {f^{jk}}_i I_k(\sigma') \right) \delta(\sigma-\sigma')-\delta_{i}^j\delta'(\sigma-\sigma') \label{IKcur}
\end{eqnarray}
 is a Poisson-Lie sigma-model.
The Hamiltonian vector fields $ 
X_{K^{i}} \, \cdot  := \{  \cdot, K^{i} \} \nonumber %\label{hamv}
$  associated to the coordinates functions $K^{i}$ for $T^{*}SU(2)$ close a non-Abelian algebra according to the following:
%Since  $K^{i}, I_{i}$ are coordinate functions for the target phase space  $T^{*}SU(2)$,  $SU(2)\ltimes \mathfrak{\tilde{g}}$,  
%with $K^{i}$ base coordinates and $I_{i}$ fiber coordinates, we associate to $K^i$ the Hamiltonian vector fields
\begin{equation}
[X_{K^{i}}, X_{K^{j}} ] =  X_{\{ K^{i}, K^j \}} = i \tau f^{ij}_{k} X_{K^{k}}  \,\,\,   \nonumber
\end{equation}
%spanning the fibers which are isomorphic to  the vector space $\R^3$
from which one can see that, because of the non-trivial Poisson bracket  in eq. (\ref{KKcur}), % $ \{K^i(\sigma),K^j(\sigma')\}=  i \tau{f^{ij}}_kK^k(\sigma') \delta(\sigma-\sigma')$,
 the structure constants of the dual Lie algebra $\mathfrak{sb}(2, \mathbb{C})$ appear and  in the limit $\tau\rightarrow 0$  the Abelian structure of the starting model over $T^*SU(2)$ is recovered. 
 
A dual formulation of this property can be given in terms of the Hamiltonian vector fields associated with the currents $I_i$, say $X_{i}$, that can be seen to close the Lie  algebra $\mathfrak{su}(2)$.
%Being  $K^{i}, I_{i}$  coordinate functions for the target phase space of the model,  %$SU(2)\ltimes \mathfrak{\tilde{g}}$, 
% with $K^{i}$ base coordinates and $I_{i}$ fiber coordinates,  Hamiltonian vector fields  associated to $K^{i}$, 
%\be
%X_{K^{i}} := \{  \cdot, K^{i} \} \label{hamv}
%\ee
%spanning the fibers which are isomorphic to   the vector space $\R^3$, close a non-Abelian algebra:
%\begin{equation}
%[X_{K^{i}}, X_{K^{j}} ] =  X_{\{ K^{i}, K^j \}} = i \tau f^{ij}_{k} X_{K^{k}}  \,\,\, .
%\end{equation}
%\pause
%\item
%The dual Lie algebra $\mathfrak{sb}(2, \C)$ is obtained, and  in the limit $\tau\rightarrow 0$  the Abelian structure of the starting model over $T^*SU(2)$ is recovered. 
%\pause
%\item
%A dual formulation of this property can be given in terms of the Hamiltonian vector fields associated with the currents $I_i$. 
 Hence, they can be regarded  as one-forms over the dual Lie algebra, which has become non-Abelian. We have then:
\begin{equation}
dX_i ( \tilde{X}^{j}, \tilde{X}^{k}) = - X_{i} ( [\tilde{X}^{j}, \tilde{X}^{k}]) = - f_{i}^{\,\, jk} 
\end{equation}
reproducing, in this way, the commonly used definition of Poisson-Lie structure.

\subsection{Poisson-Lie dual models}
From the Hamiltonian formulation of the $SU(2)$ chiral model it has been seen that it is possible to describe the dynamics in terms of the centrally extended current algebra $\mathfrak{c}_2 = \mathfrak{sl}(2,\mathbb{C})(\mathbb{R})$. Therefore one can look for a model whose target space is the dual group of $SU(2)$.

At this aim, let us consider the Poisson algebra $\mathfrak{c}_2$ (with central extension), represented by Eqs. (\ref{IIcur}), (\ref{KKcur}) and (\ref{IKcur}) and let us introduce another imaginary parameter $\alpha$ in such a way to make the role of the  subalgebras $\mathfrak{su}(2)(\mathbb{R})$ and $\mathfrak{sb}(2,\mathbb{C})(\mathbb{R})$ symmetric:
\begin{eqnarray}
\{I_i(\sigma),I_j(\sigma')\} &= & i\alpha\;  {\epsilon_{ij}}^k I_k(\sigma)\delta(\sigma-\sigma')  \nonumber%\label{IIcural} 
\\
\{K^i(\sigma),K^j(\sigma')\}&=&  i \tau{f^{ij}}_kK^k(\sigma') \delta(\sigma-\sigma') \nonumber %\label{KKcural}
\\
\{I_i(\sigma),K^j(\sigma')\} &=  &\!\!\! \left(i\alpha  K^k(\sigma'){\epsilon_{ki}}^j+i \tau   {f^{jk}}_i I_k(\sigma') \right) \delta(\sigma-\sigma')-\delta_{i}^j\delta'(\sigma-\sigma')   \label{IKcural}
\end{eqnarray}
In  the limit $i \tau\rightarrow 0$, the semi-direct sum $\mathfrak{su}(2)(\mathbb{R})\ltimes \mathfrak{a}$ is reproduced, while the limit $i\alpha\rightarrow 0$ yields $ \mathfrak{sb}(2,\mathbb{C})(\mathbb{R})\ltimes \mathfrak{a}$. 
For all non zero values of the two parameters,  the algebra is isomorphic to $\mathfrak{c}_2$, and,  by suitably rescaling the fields, one gets a two-parameter family of models, all equivalent to the Principal Chiral Model. 

Since the role of $I$ and $K$ is now symmetric, one can perform an $O(3,3)$  transformation which exchanges the momenta ${I_i}$ with the  fields  ${K^i}$, thus obtaining a new two-parameter family of models,  {\it dual} to the PCM. 
The $O(3,3)$ transformation 
\begin{equation}
\tilde{K}(\sigma)={I(\sigma)}, \;\;\; \tilde{I}(\sigma)= K(\sigma)   \nonumber %\label{tildeuntilde}
\end{equation}
yields the new Hamiltonian:
\begin{equation}
\tilde H_{\tau}=\frac{1}{2}  \int_\mathbb{R}  d\sigma  \ \tilde{I}_I \ (\tilde{ \mathcal{H}}_\tau^{-1})^{IJ}\ \tilde{I}_J,   \nonumber
\end{equation}
with 
\begin{equation} 
\begin{matrix}
{\tilde{\mathcal H}}_{\tau}= \left(
\begin{array}{cc}{h}^{\,ij} ({\tau}) & i \tau \epsilon^{il3} \delta_{lj}  \\
i\tau \delta_{il}\epsilon^{jl3} &\delta_{ij} \
\end{array} 
\right) 
\end{matrix}     
\end{equation}  
%\begin{equation}
%\tilde{H}_{\tau}= \frac{1}{2}  \int_\mathbb{R} d\sigma \  \left[\tilde{K} _s h^{sl}(\tau) \tilde{K}_l  + \tilde{I}^i\delta_{ij} \tilde{I}^j + 2 i\tau {\epsilon}^{sl3} \tilde{K}_s \tilde{I}^q \delta_{l q}\right]  \nonumber  
%\end{equation}
and $ \tilde{I}_J = ( \tilde{I}^j, \tilde{K}_j ) $ and 
with Poisson algebra:
\begin{eqnarray}
\!\!\!\!\!\{\tilde{K}_i(\sigma),\tilde{K}_j(\sigma')\}&= & i\alpha\;  {\epsilon_{ij}}^k \tilde{K}_k(\sigma)\delta(\sigma-\sigma') \nonumber \\
\!\!\!\!\!\{\tilde{I}^i(\sigma),\tilde{I}^j(\sigma')\}&=&  i \tau{f^{ij}}_k \tilde{I}^k(\sigma') \delta(\sigma-\sigma') \nonumber 
\end{eqnarray}
\begin{equation}
\!\!\!\!\!\{\tilde{K}_i(\sigma),\tilde{I}^j(\sigma')\} = \left(i\alpha  \tilde{I}^k(\sigma'){\epsilon_{ki}}^j+i \tau  {f^{jk}}_i \tilde{K}_k(\sigma') \right) \delta(\sigma-\sigma')-\delta_{i}^j\delta'(\sigma-\sigma') \,\,\, . \nonumber
\end{equation}
The metric is Riemannian, with determinant equal to 1 and such that
\begin{equation}
{\tilde{\mathcal H}}^T_{\tau}\eta {\tilde{\mathcal H}}_{\tau}=\eta.  \nonumber
\end{equation}
The  Poisson algebra implies that the new family of models, {\em Dual Principal Chiral Models}, have  as target configuration space the group manifold of $SB(2,\mathbb{C})$, spanned by the fields $\tilde{K}_i$, and momenta $\tilde{I}^i$ which span the fibers of the target phase space.
%which, for the ridefinition \eqn{tilJ} becomes
%\beqa
%\{\tJ_i(\sigma),\tJ_j(\sigma')\}&= & \tJ_k\left( i\alpha\;  {\epsilon_{ij}}^k-\frac{2\tau^2}{1-\tau^2} {f^{kn}}_j \epsilon_{jn3}\right) \delta(\sigma-\sigma') \label{tJtJcur }\\
%\{\tI^i(\sigma),\tI^j(\sigma')\}&=&  i \tau{f^{ij}}_k \tI^k(\sigma') \delta(\sigma-\sigma')\label{tItIcur}\\
%\{\tJ_i(\sigma),\tI^j(\sigma')\} &=&\left(i\alpha  \tI^k(\sigma'){\epsilon_{ki}}^j+i \tau   {f_i}^{jk}\tK_k(\sigma') \right) \delta(\sigma-\sigma')-\delta_{i}^j\delta'(\sigma-\sigma') \label{tItKcur}
%\eeqa
%In strict analogy with what found,  the analysis performed can be repeated step by step $ \rightarrow$ 
The DPCM are Poisson-Lie sigma models. Moreover, the two families are dual to each other by construction.

 It is  natural, within the Lagrangian approach,  to introduce  the fields $\tilde g:(t,\sigma)\rightarrow SB(2,\mathbb{C})$ and one-forms valued in the Lie algebra $\mathfrak{sb}(2,\mathbb{C})$, in terms of which a natural Lagrangian can be defined on the Lie-Poisson dual to $SU(2)$.  The Hamiltonian will then be  obtained by Legendre transform, together with a Poisson algebra which will result to be  isomorphic to $\mathfrak{c}_3=\mathfrak{sb}(2,\mathbb{C})(\mathbb{R})\ltimes \mathfrak{a}$. This new Hamiltonian will be related to the two-parameter family of dual models introduced above, through a $B$-transformation.
 
 The details of this procedure are reported in ref. \cite{MPV2019}. In the following the main results are summarized.
 
 The action of the proposed model is a straightforward extension  of the one in eq. (\ref{startingac})  to fields $\tilde\phi : (t,\sigma) \in  \mathbb{R}^{1,1}\rightarrow \tilde{g}\in   SB(2,\mathbb{C})$, with  Lie algebra valued  left-invariant one-forms $\tilde{g}^{-1}\mathrm{d}\tilde g$ whose pull-back to $\mathbb{R}^{1,1}$ is given by:
\begin{equation}
\tilde\phi^*(\tilde{g}^{-1}\mathrm{d}\tilde g)= (\tilde g^{-1}\partial_{t}\tilde g )_i \tilde e^i \,\mathrm{d}t+(\tilde g^{-1}\partial_{\sigma}\tilde g)_i  \tilde e^i\,\mathrm{d}\sigma.
\end{equation}
We have then:
\begin{equation}\label{dualchiact}
\tilde S=\frac{1}{2}\int_{\mathbb{R}^{1,1}} {\mathcal Tr} \left[ \phi^*(\tilde g^{-1}d\tilde g) \wedge %\h 
*\phi^*( \tilde g^{-1}d \tilde g)\right],
\end{equation}
where ${\mathcal Tr} $ stands for the non-degenerate product in the Lie algebra $\mathfrak{sb}(2,\C)$,  given by eq. (\ref{2.30}) . % and the Hodge star operator acts as $* \dd t=\dd\sigma, *\dd\sigma=\dd t$, yielding

It can be rewritten as:
\begin{equation}
\tilde S= \frac{1}{2}\int_{\mathbb{R}^{1,1}}dt d\sigma\ \bigl[  (\tilde g^{-1}\partial_t\tilde{g})_i  (\tilde g^{-1}\partial_t\tilde{g})_j - (\tilde{g}^{-1}\partial_{\sigma}g)_i  (\tilde{g}^{-1}\partial_{\sigma}g)_j   \bigr]h^{ij} . \label{actdu}
\end{equation}
The action functional is invariant under left $SB(2,\C)$ action.
 The Euler-Lagrange equations
\begin{equation}
h^{ij}\left(\partial_t(\tg^{-1}\partial_t \tg)_j-\partial_{\sigma}(\tg^{-1}\partial_{\sigma}\tg)_j\right)= {\sf L}_{\tX^i} \tilde L
% ={f^i}_{jk} \left(..\right).  
\end{equation}
%\textcolor{blue}{[COMMENT TO BE REMOVED: The last term replaces in this case $\frac{\del \tilde L}{\del \phi^i}$, $\phi^i$ being here the fields $g(\sigma)$]}
with $\tX^i(\sigma) $ the left-invariant vector fields over the group manifold and $\tilde L$ the Lagrangian, 
may be rewritten in terms of  an equivalent system of two first order partial differential equations, introducing, as for the $SU(2)$ PCM,  the currents\footnote{No factor  two is needed here because ${\mathcal Tr}(\te^i\te^j)=\delta^{ij}$}:
\begin{equation}
\tilde A_i=(\tg^{-1}\partial_t \tg)_i , \quad {\tJ}_i=(\tg^{-1}\partial_{\sigma}\tg)_i .
\label{currdu}
\end{equation}
The  Lagrangian becomes then:
\be
\tilde L= \frac{1}{2}\int_\R \dd\sigma (\tilde A_ih^{ij} \tilde A_j - \tJ_ih^{ij} \tilde J_k) \label{unmodualag}
\ee
and the equations of motion read:
\begin{eqnarray} 
h^{ij}(\partial_t \tilde A_j- \partial_{\sigma}\tJ_j)  & = &{ f^{si}}_{l} h^{lj} (\tA_s \tA_j -\tJ_s\tJ_j),  
\label{equivdu1} \\
\partial_t \tJ & = & \partial_{\sigma} \tilde A  -[\tilde A,\tJ],
 \label{equivdu2}
\end{eqnarray}
being the latter a condition for the  existence of  $\tg \in SB(2,\C)$ that admits the expression of the currents in the form given by eq. (\ref{currdu}). 
At fixed $t$, all elements $\tg$ satisfying the boundary condition $ \lim_{\sigma\to\pm\infty} \tg(\sigma)=1$
 form the  infinite-dimensional Lie group $SB(2,\C)({\R})\equiv \mathrm{Map}(\R,SB(2,\C))$, given by smooth maps $\tg: \sigma\in \mathbb{R}\rightarrow \tg(\sigma)\in SB(2,\C)$ which are constant at infinity. 
 
 At fixed time, the currents $\tJ$ and $\tilde A$ take values in the Lie algebra $\mathfrak{sb}(2,\C)(\mathbb{R})$ of functions from $\mathbb{R}$ to $\mathfrak{sb}(2,\C)$ that are sufficiently fast decreasing at infinity to be square-integrable.  Therefore the tangent bundle description of the dual dynamics can be given in terms of $(\tJ,\tilde A)$, with $\tilde A$ the left generalized velocities, while $\tJ$ playing the role of left configuration space coordinates.

By introducing left momenta
\be
\tI^i=\frac{\delta \tilde L}{\delta \ (\tg^{-1}\partial_t \tg)_i}=(\tg^{-1}\partial_t\tg)_jh^{ij}={\tilde A}_jh^{ij}
\end{equation}
and inverting for the generalized velocities, one obtains the Hamiltonian:
\be 
\tilde H=\frac{1}{2}\int_{\mathbb{R}} \mathrm{d}\sigma \left( \tI^i \tI^j  h_{ij}+\tJ_i \tJ_j  h^{ij} \right) = \frac{1}{2}\int_{\mathbb{R}} \mathrm{d}\sigma \tI_I ({{\tilde{\mathcal{K}}}_0}^{-1})^{IJ} \tI_J   \label{sbham}
\ee
with 
\be \label{metsb0}
{\tilde{\mathcal K}}_0= \left(
\begin{array}{cc}
 { h}^{\,ij}  & 0  \\
 0 &  h_{ij}
\end{array} \right)
\ee
and $\tI_J=(\tI^j,\tJ_j)$,  while the equal-time 
Poisson brackets can be derived in the usual way  from the action functional; they result to be
\begin{eqnarray} 
\{\tilde{I}^i(\sigma),\tilde{I}^j(\sigma')\} &= & {f^{ij}}_k \tilde{I}^k(\sigma)\delta(\sigma-\sigma'),  \label{poidu1} \\
\{\tilde{I}^i(\sigma),\tilde{J}_j(\sigma')\} &= &\tilde{J}_k(\sigma) {f^{ki}}_j \delta(\sigma-\sigma') -\delta^i_{j}\delta'(\sigma-\sigma'),  \label{poidu2} \\
\{\tilde{J}_i(\sigma),{\tJ}_j(\sigma')\}& = & 0  \label{poidu3}
\end{eqnarray}
yielding the equations of motion
\beqa
\del_t \tilde{I}^i(\sigma)&=&  \tI^s\tI^r {f^{ji}}_s h_{rj}  - \tJ_r \tJ_s {f^{si}}_j h^{rj}  + \delta^i_j\, h^{rj}\, \del_\sigma \tJ_r  \\
\del_t \tJ_i (\sigma)&=& \left(\tI^s \tJ_k {f^{kj}}_i + \delta^j_i \del_\sigma \tI^s\right)h_{sj} \,\,\, . 
\eeqa
The Poisson brackets \eqn{poidu1}-\eqn{poidu3} realize the current algebra 
 $\mathfrak{c_3}=\mathfrak{sb}(2,\mathbb{C})(\mathbb{R})\ltimes\mathfrak{a}$, which we have already regarded as the limit $i\alpha\rightarrow 0$ of the algebra \eqn{IIcur}-\eqn{IKcur}.
 %\footnote{Let us notice that, upon defining $\tilde W_i= \tJ_i - \frac{1}{2}{\epsilon}_{iq3}\tI^q$ 
  %we have an equivalent description in terms of the Hamiltonian 
  %\be 
  %\tilde H=\frac{1}{2} \int_\R\dd\sigma \  \left[\tilde W _i h^{ij} \tilde W_j  + \tilde I^ih_{ij} \tilde I^j \right]
  %\ee and Poisson algebra 
 %\beqa
%\{{\tilde W}_i(\sigma),\tilde W_j(\sigma')\}&= &{f^{kn}}_j \epsilon_{in3}\tilde W_k(\sigma)\delta(\sigma-\sigma') \label{defo1}\\
%\{\tI^i(\sigma),\tI^j(\sigma')\}&=& {f^{ij}}_k \tI^k(\sigma') \delta(\sigma-\sigma')\label{defo2} \\
%\{{\tilde W}_i(\sigma),\tI^j(\sigma')\} &=& \left( -\tI^n(\sigma')\epsilon_{kn 3} {f^{jk}}_i +  {f^{jk}}_i \tilde W_k(\sigma') \right) \delta(\sigma-\sigma')-\delta_{i}^j\delta'(\sigma-\sigma') \label{defo3}
%\eeqa
 %\label{note}
%}
 %, upon identifying $\tilde J$ with $\tilde K$.  

Similarly to the $SU(2)$ PCM,   the currents $(\tJ, \tI)$ may be identified with the cotangent space left coordinates for $T^*SB(2,\C)(\R)$. 
However, differently from  $T^*SU(2)$, $T^*SB(2,\C)$ is not symplectomorphic to $SL(2,\C)$, the two spaces being topologically different to start with.   Therefore, certainly the model cannot be given an equivalent description in terms of an $SL(2,\C)(\R)$ algebra. 
%Indeed,  it will be shown, in the next section, that  the  $SB(2,\C)$ PCM Hamiltonian obtained here through Legendre transform can be related to the DPCM models previously found,   through a $B$-transformation, but not its Poisson algebra.   %We claim therefore that the two models are dual to each other.  

\section{The Doubled Principal Chiral Model}
\subsection{The Lagrangian formalism}
Following the same spirit that led to the string doubled world-sheet action in eq. (\ref{Tseytlinact}), one can construct a Doubled Principal Chiral Model, i.e. a $SL(2, \mathbb{C})$ Principal Chiral Model.  At this aim, the ingradients the  for getting the Lagrangian are the group valued field $ \Phi: \mathbb{R}^{1,1} \rightarrow \gamma  \in  SL(2,\mathbb{C})$
and the left-invariant  Maurer-Cartan one-form $\gamma^{-1}\mathrm{d}\gamma  \in \Omega^1(SL(2,\mathbb{C})) \otimes {\mathfrak sl(2,C)} $, pulled back to $\mathbb{R}^{1,1} $:
%with pull-back to to $\mathbb{R}^{1,1} $:
\begin{equation} %\label{mauca}
\Phi^*( \gamma^{-1}\mathrm{d}\gamma)=\gamma^{-1}\partial_t \gamma \mathrm{d}t+\gamma^{-1}\partial_{\sigma}\gamma \mathrm{d}\sigma  \nonumber
\end{equation}
By using the Lie algebra basis $e_I=(e_i, \tilde{e}^i)$ one has:
\begin{equation}
 \gamma^{-1}\partial_t \gamma =  \dot{ \mathbf{Q}}^I e_I,  \,\,\,; \,\,\,
 \gamma^{-1}\partial_{\sigma}\gamma =  {\mathbf{Q}'^I}e_I.  \nonumber 
 \end{equation}
 $\dot{ \mathbf{Q}}^I, \mathbf{Q}'^I$, left generalized coordinates, respectively given by:
\begin{equation}
 \dot{\mathbf{Q}}^I={\mbox Tr}\left( \gamma^{-1}\partial_t \gamma e_I\right),\;\;\;    {\mathbf{Q}'^I}={\mbox Tr}\left( \gamma^{-1}\partial_\sigma \gamma e_I\right)   \nonumber
 \end{equation}
 with ${\mbox Tr}$ the Cartan-Killing metric of $\mathfrak{sl}(2,\mathbb{C})$.
Then, the Lagrangian density can be rewritten in terms of the left generalized coordinates
$\dot {\bf Q}^I$ as follows:
\begin{equation}  \label{lagrdoubl}
{\mathbf{L}}=   \frac{1}{2} % \int_\R  d\sigma\,
  (k ~\eta+ \mathcal{H} )_{IJ} \left(\dot {\mathbf{Q}}^I  \dot { \mathbf{Q}}^J   -  {\mathbf{Q}'^I } { \mathbf{Q}'^J} \right)  
 \end{equation}  
 with 
\begin{equation} 
(k  \eta+ \mathcal{H} )_{IJ}
 = \left(
\begin{array}{cc} 
 \delta_{ij} & k \delta_i^j+  \epsilon_i^{\,j3}  \\
 k \delta^i_j-\epsilon^i_{\,j3} &  \delta^{ij}+ \epsilon^i_{k3} \epsilon^j_{l3} \delta^{kl} 
\end{array}
\right)
\end{equation}
 $\eta$ (Lorentzian) and $\mathcal{H}$ (Riemannian) are the left-invariant metrics on $SL(2, \mathbb{C})$ induced, respectively, by the pairings $2\mathrm{Im} \mathrm{Tr}()$ and $2\mathrm{Re}  \mathrm{Tr}()$ on $\mathfrak{sl}(2, \mathbb{C})$. They are two of the structures defining a Born geometry on $SL(2, \mathbb{C}).$ 

  The degrees of freedom are doubled. Performing a gauging of its symmetry both the Lagrangian models, with $SU(2)$ and $SB(2,C)$ target configuration spaces, can be retrieved \cite{MPV2019}.

\subsection{The Hamiltonian formalism} \label{hamifor}
 The Hamiltonian model will be interpreted as a model over the cotangent space $T^*SL(2,\C)(\R)$.  In order to obtain the Hamiltonian of the system,  the canonical momentum is computed:
\begin{equation} \label{mompi}
\bI_I=(I_i, \tilde{I}^i)=\frac{\delta\mathbf{L}}{\delta\dot{\bQ}^I}=(k\ \eta+\mathcal{H})_{IJ}\dot{\bQ}^J.  
\end{equation}
Let us recall that the matrix $(k \ \eta+ \mathcal{H})_{IJ}$ is invertible for $k^2 \neq 1$ and its inverse is
\be
[( k\, \eta + \mathcal{H})^{-1}]^{IJ}= \frac{1}{2} (1-k^2)^{-1} 
\left( \begin{array}{cc}
\delta^{ij}+ \epsilon^i_{l3} \epsilon^j_{k3}\delta^{lk}& -{\epsilon^i}_{j3}-k \delta^i_j
\\
{\epsilon_i}^{j3}-k \delta_i^j& \delta_{ij} \,\,\, 
\end{array} \right)  \,\,.    \nonumber
\ee
Therefore, the Legendre transform of the Lagrangian density in eq. (\ref{lagrdoubl}), obtained by inverting (\ref{mompi}), gives:
\begin{equation}
{\bf H}=\frac{1}{2}\int_{\mathbb{R}} \mathrm{d}\sigma \ \bigl([(k\, \eta+ \mathcal{H})^{-1}]^{IJ}\bI_I\bI_J+(k\, \eta+ \mathcal{H})_{IJ}{\bJ^I} {\bJ^J}\bigr).
\end{equation}
whereas one has for the Poisson brackets
\beqa 
\{\bI_I(\sigma'), \bI_J(\sigma'')\}&=& {C_{IJ}}^K \bI_K \delta(\sigma'-\sigma'') \label{poidoub1}\\
\{\bI_I(\sigma'), \bJ^J(\sigma'')\}&=& {C_{KI}}^J \bJ^K \delta(\sigma'-\sigma'')-\delta_I^J \delta'(\sigma'-\sigma'') \label{poidoub2}\\
\{\bJ^I(\sigma'), \bJ^J(\sigma'')\}&=& 0 \label{poidoub3}
\eeqa
by renaming $\bQ'^I\rightarrow \bJ^I$. The equations of motion read then as:
\beqa
\dot\bI_J&=&  \left\{\bI_M [ k\, \eta+ \mathcal{H})^{-1}]^{LK}\  \bI_K -  \bJ^L[ (k\, \eta+ \mathcal{H})^{-1}]_{LK}\  \bJ^K\right\} {C_{JL} }^M\nn\\
&+& \del_\sigma \bJ^L [(k\, \eta+ \mathcal{H})^{-1}]_{LJ} 
\eeqa

\section{A further generalization: the non-linear sigma model with WZ term}

The Principal Chiral Model analyzed so far can be further extended into a more general WZW model on $SU(2)$ by introducing a topological term. In so doing, one  obtains the non-linear sigma model defined by the following action:
\begin{equation}
\label{wzwaction}
S=\frac{1}{4 \lambda^2} \int_{\Sigma} Tr\left[ \varphi^*\left(g^{-1} dg\right) \wedge * \varphi^*\left(g^{-1} dg \right)\right]+\kappa S_{WZ},
\end{equation}
where $S_{WZ}$ is the so-called \textit{Wess-Zumino term}, and it is defined as
\begin{equation}
\label{wzdef}
S_{WZ}=\frac{1}{24 \pi}\int_{\mathcal{B}} Tr \left[ \tilde{\varphi^*}\left(\tilde{g}^{-1} d\tilde{g} \wedge \tilde{g}^{-1} d\tilde{g} \wedge \tilde{g}^{-1} d\tilde{g} \right)\right],
\end{equation}
where $\mathcal{B}$ is a three-manifold whose boundary is the compactification of the original two-dimensional spacetime, while $\tilde{g}$ and $\tilde{\varphi}$ are the extensions on the three-manifold $\mathcal{B}$. It is possible to show by simple homotopy arguments that such an extension always exists. 

It is well-known that there is an ambiguity in the extension of the $g$ fields to the manifold $\mathcal{B}$, i.e. there may be many ways to do such an extension, or equivalently, there may be many three-manifolds with the same boundary. However, it is straightforward to check that the variation of the WZW action remains the same up to a constant term, so classically there is no problem. However, in the quantum theory this may be a problem, indeed it is needed to make the partition function well-defined (single-valued). For compact Lie groups this requirement follows in the restriction of $\kappa$ to integer values ($\kappa$  is called the level of the theory), while for non-compact Lie groups no such a quantization condition holds.

Note that although the WZ term is expressed as a three-dimensional integral, under the variation $g \rightarrow g+ \delta g$ it gets reduced to a boundary term, which is exactly an integral over $\Sigma$ since the variation of its Lagrangian density can be written as a total derivative and then reduced on the boundary by using Stokes theorem $\int_{\mathcal{B}} d^3y \, \epsilon^{\alpha \beta \gamma} \partial_{\gamma}\left(\cdots \right)=\int_{\Sigma}d^2 \sigma \epsilon^{\alpha \beta} \left(\cdots \right)$.

The WZW model is a very important model and it is involved in several interesting applications, both theoretical and phenomenological. In string theory, in particular, it can describe strings propagating on a Lie group manifold, which are especially relevant backgrounds since they are Ricci flat and also represent the appropriate setting to analyze T-duality generalizations. For example, notice that the WZW model on $SL(2,\mathbb{R})$ has been used to construct bosonic string theories on AdS3 \cite{MO1, MO2, MO3}, or superstring theories on $AdS_3 \times S^3$ geometries can be described by the WZW on $PSU(1,1|2)$ \cite{gotz}. Furthermore, WZW models having a non semi-simple group as target space have attracted some interest in the scientific community, being the latter particularly relevant as string backgrounds \cite{nappi, kehagias}.

The equations of motion for the theory are given by
\begin{equation}
\partial_{\alpha}i^{\alpha}+\frac{\kappa \lambda^2}{4 \pi}\epsilon^{\alpha \beta} i_{\alpha}i_{\beta}=0,
\end{equation}
having defined the currents $i_{\alpha}=g^{-1}\partial_{\alpha} g$ ($\alpha=0,1$).

Even in this case, the Euler-Lagrange equations can be written as a pair of first-order partial differential equations
\begin{equation}
\partial_{t}A-\partial_{\sigma}J=-\frac{\kappa \lambda^2}{4 \pi}\left[A,J \right],
\end{equation}
\begin{equation}
\label{intcond}
\partial_t J-\partial_{\sigma}A=-\left[A,J \right],
\end{equation}
where the $A$ and $J$ currents are $\mathfrak{su}(2)$-valued and can be written as $A=\left(g^{-1}\partial_t g \right)^ie_i=A^i e_i$, $J=\left(g^{-1}\partial_{\sigma} g \right)^ie_i=J^i e_i$. As for the PCM, if we also impose the boundary condition 
\begin{equation}
\lim_{|\sigma| \rightarrow \infty} g(\sigma)=1
\end{equation}
the solution for $g$ is unique. 

In the Hamiltonian description, the model can be described by the following pair of Poisson bracket and Hamiltonian:
\begin{equation}
\begin{aligned}
{} & \{I_i (\sigma), I_j(\sigma')\}=2\lambda^2{\epsilon_{ij}}^k I_k(\sigma)\delta(\sigma-\sigma')+\frac{\kappa \lambda^4}{2 \pi} \epsilon_{ijk}J^k(\sigma)\delta(\sigma-\sigma') \\ &
\{I_i (\sigma), J^j(\sigma')\}=2\lambda^2\left[{\epsilon_{ki}}^j J^k(\sigma)\delta(\sigma-\sigma')-\delta_i^j \delta'(\sigma-\sigma') \right] \\ &
\{J^i(\sigma),J^j(\sigma') \}=0,
\end{aligned}
\end{equation}
\begin{equation}
H=\frac{1}{4\lambda^2}\int d\sigma \, \left(\delta^{ij}I_i I_j+\delta_{ij}J^i J^j \right).
\end{equation}
We can see that the Poisson algebra is the semi-direct sum of an abelian algebra $\mathfrak{a}(\mathbb{R})$ and a Kac-Moody algebra associated to $SU(2)(\mathbb{R})$ and even in this case the current algebra can be deformed to a one-parameter family of fully non-Abelian algebras, in such a way that the resulting brackets, together with a one-parameter family of deformed Hamiltonians, lead to an equivalent description of the dynamics, which is again $\mathfrak{sl}(2,\mathbb{C})(\mathbb{R})$. In this case, however, since the Poisson structure is more complicated, it requires some more manipulations and rotations of the generators to show that, see ref. \cite{BPV} for details.

After deforming the algebra and some manipulations, it can be shown that the Poisson algebra of the model can be recovered in the following form in terms of the $(\mathfrak{sl}(2,\mathbb{C}), \mathfrak{su}(2), \mathfrak{sb}(2,\mathbb{C}))$ Manin triple decomposition:
\begin{equation}
\begin{aligned}
{} & \{S_i(\sigma), S_j(\sigma') \}={\epsilon_{ij}}^kS_k(\sigma)\delta(\sigma-\sigma')+C_{\tau} \delta_{ij}\delta'(\sigma-\sigma') \\ &
\{K^i(\sigma), K^j(\sigma') \}=i\tau{f^{ij}}_k K^k(\sigma)\delta(\sigma-\sigma')+C_{\tau}\tau^2(\delta^{ij}+{\epsilon_p}^{i3} \epsilon^{jp3}) \delta'(\sigma-\sigma') \\ &
\{S_i(\sigma) ,K^j(\sigma')\}=\left[{\epsilon_{ki}}^j K^k(\sigma)  +i\tau {f^{jk}}_i S_k(\sigma)  \right] \delta(\sigma-\sigma')+\left( C'_{\tau}{\delta_i}^j+i\tau C_{\tau}{\epsilon_i}^{j3} \right)\delta'(\sigma-\sigma'),
\end{aligned}
\end{equation}
and the Hamiltonian is given by
\begin{equation}
\begin{aligned}
{} H_{\tau} & =\lambda^2 \int d\sigma \, S_I \left(\mathcal{M}_{\tau} \right)^{IJ} S_J \\ & = \lambda^2 \int d\sigma \left[(m_{\tau})^{ij} S_i S_j+(m_{\tau})_{ij} K^i K^j+S_i K^j {(m_{\tau})^i}_j+S_j K^i {(m_{\tau})_i}^j \right],
\end{aligned}
\end{equation}
where $S$, $B$  and $K$ are defined as
\begin{equation}
\label{transfsl}
\begin{aligned}
{} & S_i(\sigma)= \frac{1}{\xi(1-a^2 \tau^2)}\left[ I_i(\sigma)-a \delta_{ik} J^k(\sigma) \right] \\ &
B^i(\sigma)=\frac{i \tau}{\xi (1-a^2 \tau^2)}\left[-a i \tau \delta^{ik}I_k(\sigma)-\frac{1}{i \tau} J^i(\sigma) \right],
\end{aligned}
\end{equation}
\begin{equation}
K^i(\sigma)=B^i(\sigma)-i\tau \epsilon^{i \ell 3}S_{\ell}(\sigma).
\end{equation}
In the previous expressions we identified $a=\frac{\kappa \lambda^2}{4\pi}$ and $\xi=2\lambda^2(1-\tau^2)$, while the constants $C_{\tau}$ and $C'_{\tau}$ in the central terms are defined as $C=\frac{a}{\lambda^2(1-a^2\tau^2)^2}$ and $C'=\frac{(1+a^2\tau^2)}{2 \lambda^2 (1-a^2\tau^2)^2}$.

Even in this case one may introduce another imaginary parameter $\alpha$ in order to make the role of the two subalgebras $\mathfrak{su}(2)$ and $\mathfrak{sb}(2,\mathbb{C})$ symmetric without spoiling the dynamics of the model. Again, differently from the PCM some more manipulations are required \cite{BPV}, but the final result for the two-parameter family of Poisson brackets is the following:
\begin{equation}
\begin{aligned}
{} & \{U_i(\sigma), U_j(\sigma') \}=i\alpha{\epsilon_{ij}}^k U_k(\sigma)\delta(\sigma-\sigma')-\frac{a \alpha^2}{\lambda^2 \left(1+a^2 \tau^2 \alpha^2 \right)^2} \delta_{ij}\delta'(\sigma-\sigma') \\ &
\{V^i(\sigma), V^j(\sigma') \}=i\tau{f^{ij}}_k V^k(\sigma)\delta(\sigma-\sigma')+\frac{a \tau^2}{\lambda^2 \left(1+a^2 \tau^2 \alpha^2 \right)^2}(\delta^{ij}+{\epsilon_p}^{i3} \epsilon^{jp3}) \delta'(\sigma-\sigma') \\ &
\{U_i(\sigma), V^j(\sigma')\}=\left[i\alpha{\epsilon_{ki}}^j V^k(\sigma)  +i\tau {f^{jk}}_i U_k(\sigma)  \right] \delta(\sigma-\sigma') \\ & \quad \quad \quad \quad \quad \quad \quad \quad +\frac{1}{2\lambda^2\left(1+a^2 \tau^2 \alpha^2 \right)^2}\left[ \left(1-a^2 \tau^2 \alpha^2\right){\delta_i}^j+2 a\, i\tau \, i\alpha \, {\epsilon_i}^{j3} \right]\delta'(\sigma-\sigma'),
\end{aligned}
\end{equation}
and the new generators $U$ and $V$ are obtained from the old ones as
\begin{equation}
U_i=i\alpha S_i, \quad V^i=\frac{1}{i\alpha}B^i-i\tau \epsilon^{i \ell 3} S_{\ell}.
\end{equation}
The two-parameter family of Hamiltonians, together with which the Poisson brackets still give the old dynamics, is
\begin{equation}
\begin{aligned}
{} & H_{\tau, \alpha}=\lambda^2 \int d\sigma \, U_I \left(\mathcal{M}_{\tau, \alpha} \right)^{IJ} U_J \\ & = \lambda^2 \int d\sigma \left[(m_{\tau, \alpha})^{ij} U_i U_j+(m_{\tau, \alpha})_{ij} V^i V^j+U_i V^j {(m_{\tau, \alpha})^i}_j+U_j V^i {(m_{\tau, \alpha})_i}^j \right],
\end{aligned}
\end{equation}
with 
\begin{equation}
\mathcal{M}_{\tau, \alpha}=\begin{pmatrix}
  \frac{1+a^2 \tau^4 \alpha^4}{(i\alpha)^2}\delta^{ij}-\tau^2(1+a^2){\epsilon_p}^{i3} \epsilon^{pj3} & i\tau i\alpha(1+a^2){\epsilon_j}^{i3}-a(1-\tau^2\alpha^2){\delta^i}_j\\
  i\tau i\alpha (1+a^2){\epsilon_i}^{j3}-a(1-\tau^2\alpha^2){\delta_i}^j & (i\alpha)^2(1+a^2)\delta_{ij}
 \end{pmatrix}.
\end{equation}
Now, since the role of $U$ and $V$ is symmetric, by exchanging the momenta $U$ with the fields $V$ one obtains a new two-parameter family of models, which consists of \textit{dual} models. In particular it can be checked that the transformation
\begin{equation}
\tilde{V}(\sigma)=U(\sigma), \quad \tilde{U}(\sigma)= V(\sigma)
\end{equation}
is an $O(3,3)$ rotation. 

Explicitly, under such a rotation one obtains the dual Hamiltonians
\begin{equation}
\tilde{H}_{\tau, \alpha}=\lambda^2 \int d\sigma \left[(m_{\tau, \alpha})^{ij} \tilde{V}_i \tilde{V}_j+(m_{\tau, \alpha})_{ij} \tilde{U}^i \tilde{U}^j+\tilde{V}_i \tilde{U}^j {(m_{\tau, \alpha})^i}_j+\tilde{V}_j \tilde{U}^i {(m_{\tau, \alpha})_i}^j \right],
\end{equation}
and the dual Poisson algebras
\begin{equation}
\begin{aligned}
{} & \{\tilde{V}_i(\sigma), \tilde{V}_j(\sigma') \}=i\alpha{\epsilon_{ij}}^k \tilde{V}_k(\sigma)\delta(\sigma-\sigma')-\frac{a \alpha^2}{\lambda^2 \left(1+a^2 \tau^2 \alpha^2 \right)^2} \delta_{ij}\delta'(\sigma-\sigma') \\ &
\{\tilde{U}^i(\sigma), \tilde{U}^j(\sigma') \}=i\tau{f^{ij}}_k \tilde{U}^k(\sigma)\delta(\sigma-\sigma')+\frac{a \tau^2}{\lambda^2 \left(1+a^2 \tau^2 \alpha^2 \right)^2}(\delta^{ij}+{\epsilon_p}^{i3} \epsilon^{jp3}) \delta'(\sigma-\sigma') \\ &
\{\tilde{V}_i(\sigma), \tilde{U}^j(\sigma')\}=\left[i\alpha{\epsilon_{ki}}^j \tilde{U}^k(\sigma)  +i\tau {f^{jk}}_i \tilde{V}_k(\sigma)  \right] \delta(\sigma-\sigma') \\ & \quad \quad \quad \quad \quad \quad \quad \quad +\frac{1}{2\lambda^2\left(1+a^2 \tau^2 \alpha^2 \right)^2}\left[ \left(1-a^2 \tau^2 \alpha^2\right){\delta_i}^j+2 a\, i\tau \, i\alpha \, {\epsilon_i}^{j3} \right]\delta'(\sigma-\sigma'),
\end{aligned}
\end{equation}
which make it clear that this new two-parameter family of models has target configuration space the group manifold of $SB(2,\mathbb{C})$, spanned by the fields $\tilde{V}_i$, while momenta $\tilde{U}^i$ span the fibers of the target phase space. Hence, this family effectively consists of dual models. 

Note that a natural dual WZW model having $SB(2,\mathbb{C})$ as target space from the very beginning can be formulated, but it is more difficult than the PCM situation, due to the presence of the WZ term and the fact that $SB(2,\mathbb{C})$ is not a semi-simple Lie group. In fact, one can show that the natural metric $h$ on $SB(2,\mathbb{C})$ extracted from the generalized one on $SL(2,\mathbb{C})$ in eq. (\ref{sb2cmetric}) cannot be used and a further generalization has to be taken into account \cite{BPV}. 

Furthermore, while for the PCM case the gauging of the global isometries of the doubled model is straightforward due to the fact that minimal coupling is enough to obtain a gauge-invariant action, the same can not hold for the WZW model, for which the minimal coupling gauged pull back of the $3$-form which is integrated in the WZ term may no longer be closed. There are conditions under which a gauging of the WZW model on a semi-simple Lie group with the natural symmetric Cartan-Killing form can be performed, but it is not clear what conditions are required in our doubled WZW case in which the product used cannot be the natural Cartan-Killing one.

\section{Conclusion and Outlook}

 It has been shown here that the $SU(2)$ Principal Chiral Model, in its Hamiltonian formulation, can be given an equivalent description in terms of currents which span  a target phase space isomorphic to the group manifold of $SL(2,\C)$. Their Poisson algebra can be given the structure of the centrally extended  affine algebra $\mathfrak{sl}(2,\C)(\R)$.  %Following a previous paper of the authors, \cite{MPV18}, the model is here studied  as a higher dimensional generalization of the Isotropic Rigid Rotor dynamics  with the aim of further deepening its remarkable geometric structures. 
\\
The standard Hamiltonian formulation of the $SU(2)$ PCM model %
is defined on the target phase space  $T^*SU(2) \simeq SU(2) \ltimes \mathbb{R}^3$  that, 
%exploits the fact that the dynamics is fully described by fields, the currents, which  span $T^*SU(2)$ as target phase space  and act as infinitesimal generators of an affine algebra which is the semi-direct sum  $\mathfrak{su}(2)(\R)\dot\oplus \mathfrak{a} (\R)$. 
%$T^*SU(2)$, 
as a Lie group, is  the trivial Drinfel'd double of the group $SU(2)$,  the classical double. The latter  gives rise to a fully nontrivial Drinfel'd double, the group $SL(2,\C)$, when the Abelian subalgebra of the semidirect sum is deformed to that of $SB(2,\C)$. %By exploiting this property, we first review in detail 
 A whole family of  equivalent PCM models described in terms of  current algebra of the group $SL(2,\C)$ has been reviewed, showing that they can actually be interpreted in terms of Born geometries related by B-transformations. By performing $O(3,3)$ transformations of such  a family, one finds a parametric family of T-dual PCM models, with target configuration space the group $SB(2,\C)$,   the Poisson-Lie dual  of $SU(2)$ in the Iwasawa decomposition of the Drinfel'd double $SL(2, \mathbb{C} )$.  %Poisson-Lie symmetries are discussed. 
%A relevant result obtained in this context is that the PCM, in the formulation given by one of the equivalent Hamiltonians, in particular the one in Eq. \eqn{modiha3}, together with the Poisson algebra \eqn{IIcur}-\eqn{IKcur}, is a Poisson-Lie sigma-model.
Then, a natural Lagrangian model has been constructed directly on the dual group $SB(2,\C)$.  %Its relation to the dual models previously introduced is still unclear to us and needs further analysis.   
 %we have introduced 
A double PCM has been then constructed with the group manifold of $SL(2,\C)$ as its target configuration space and $TSL(2,\C)$ as the target tangent  space. The degrees of freedom are thus doubled. Performing a gauging of its symmetries, both of the Lagrangian models, with  $SU(2)$ and $SB(2,\C)$ target configuration spaces, can be retrieved. 
 %The natural $SB(2,\C)$ PCM model  constructed has an Hamiltonian formulation given by the Hamiltonian \eqn{sbham} and the Poisson algebra \eqn{poidu1}-\eqn{poidu3}. Alternatively, it can be described %according to footnote [$\ref{note}$], 
 %by using the Hamiltonian Eq. \eqn{sbdefoham} and with the Poisson algebra represented by Eqs. \eqn{defo1}-\eqn{defo3}. 
%An interesting new kind of duality has emerged out in this context:  the models that have been obtained  by performing a T-duality transformation of target space,  namely an $O(3,3)$ rotation, are described by either of  those two particular Hamiltonian functions but with Poisson algebras exchanged. The meaning of this new duality is still under investigation.

 %Finally, in order to build a model where the symmetries exhibited by the dynamics are manifest, a parent action is constructed with target configuration space  the Drinfel'd double $SL(2,\C)$, hence doubling the degrees of freedom.   From it, either of the dual partner models can be recovered, by gauging one of its global symmetries. 

Finally, a further extension of the $SU(2)$ PCM by adding a Wess-Zumino term \cite{BPV} has been shortly illustrated. %This could provide a deeper insight, among other things,  on the geometric structures  of String Theory on $AdS_3$, the study of which is interesting from the point of view of the $AdS/CFT$ correspondence since it enables to study the correspondence beyond the gravity approximation \cite{MO, MO2,MO3}.

%Last but not least, all what we have learnt from this model could be further extended to the world-sheet string action. In this case, a manifestly $O(d,d)$-invariant action may be written, considering that the configuration space is no longer a Lie group, but a differentiable manifold.  It would be interesting to follow this way, in which $O(d,d)$-invariance is implemented writing a doubled string action, as discussed for Principal Chiral Models, and then performing the low energy limit. This limit  result should reproduce all the results so far obtained in Double Field Theory.

\vspace{.2cm}

{\bf Acknowledgments} 
The authors are indebted to Vincenzo E. Marotta and Patrizia Vitale, coauthors of the papers \cite{MPV2019} and \cite{BPV} their talks are based on. 

F.P. would like to thank the organizers of the Workshop on {\em Recent Developments in Strings and Gravity} for the invitation to give a talk as a main speaker.


\begin{thebibliography}{99}

\bibitem{GPR} 
A. Giveon,  M. Porrati and E. Rabinovici, \emph{Target Space Duality in String Theory}, Phys. Rept. {\bf 244}  (1994) 77   [{\tt hep-th/9401139}].

\bibitem{AAL}
E. Alvarez,  L. Alvarez-Gaum\'e and  Y. Lozano, \emph{An Introduction to T-Duality in String Theory}, Nucl. Phys. Proc. Suppl. {\bf 41} (1995) 1 [{\tt hep-th/94010237}].

\bibitem{D}
M. J. Duff, \emph{Duality Rotations in String Theory}, Nucl. Phys. {\bf B335} (1990) 610.

\bibitem{Plauschinn2018}
E. Plauschinn, \emph{Non-Geometric Backgrounds in String Theory}, Phys. Rept. {\bf 798} (2019) 1. 

\bibitem{Rennecke}
F. Rennecke, \emph{$O(d,d)$-Duality in String Theory}, JHEP {\bf 1410} (2014) 69 [{\tt arXiv:1404.0912 [hep-th]]}.

\bibitem{Maharana}
J. Maharana, \emph{The Worldsheet Perspective of T-duality Symmetry in String Theory}, Int. J. Mod. Phys. {\bf A28} (2013) 1330011 [{\tt  arXiv:1302.1719  [hep-th]]}.

\bibitem{Buscher1987}
T. H. Buscher, \emph{A Symmetry of the String Background Field Equations}, Phys. Lett. {\bf B194} (1987) 59.

\bibitem{Buscher1988}
T. H., Buscher, \emph{Path Integral Derivation of Quantum Duality in Nonlinear Sigma Models}, Phys. Lett. {\bf B201} (1988) 466-472.

\bibitem{Hull2005}
C. M. Hull, \emph{A Geometry for Non-Geometric String Backgrounds}, JHEP {\bf  0510}  (2005) 065 [{\tt hep-th/0406102]}.

\bibitem{Tseytlin1990}
A. A. Tseytlin, \emph{Duality Symmetric Formulation of String World Sheet Dynamics}, Phys. Lett. {\bf B242} (1990) 163.
\bibitem{Tseytlin1991}
A. A. Tseytlin, \emph{Duality Symmetric Closed String Theory and Interacting Chiral Scalars}, Nucl. Phys. {\bf B350} (1991) 395.

\bibitem{Siegel1993}
W. Siegel, \emph{Two Vierbein Formalisms for String Inspired Axionic Gravity}, Phys. Rev. {\bf D47} (1993) 5453 {\tt [hep-th/9302036]}.

\bibitem{Lee2014}
K. Lee and J.-H. Park, \emph{Covariant Action for a String in Doubled-yet-Gauged Spacetime}, Nucl. Phys. {\bf B880} (2014) 134  [{\tt arXiv:1307.8377 [hep-th]]]}.

\bibitem{Nibbelink2013}
S. Groot Nibbelink and P. Patalong, \emph{ A Lorentz Invariant Doubled World-Sheet Theory}, Phys. Rev. {\bf D 87} (2013) 041902 {[\tt arXiv:1207.6110 [hep-th]]}. 

\bibitem{Angelis2014}
L. De Angelis, G. Gionti S.J., R. Marotta, and  F. Pezzella,  \emph{Comparing Double String Theory Action}, JHEP {\bf 1404} (2014) 171 {[\tt arXiv:1312.7367 [hep-th]]}.

\bibitem{Copland2012}
N. B. Copland, \emph{A Double $\sigma$-model for Double Field Theory}, JHEP {\bf 1204} (2012) 044 {[\tt arXiv:1111.1828  [hep-th]]}. 
\bibitem{Park2013}
J. H. Park, \emph{Comments on Double Field Theory and Diffeomorphisms}, JHEP {\bf 1306} (2013) 098 {[\tt arXiv:1304.5946 [hep-th]]}. 

\bibitem{Berman2008}
D.S. Berman,  N.B. Copland, and D.C. Thompson, \emph{Background Field Equations for the Duality Symmetric String}, Nucl. Phys. {\bf B791} (2008) 175 [{\tt arXiv:0708.2267[hep-th]]}.

\bibitem{Ma2015}
Chen-Te Ma, \emph{One-loop {$\beta$} function of the double sigma model with constant background}, JHEP {\bf 1504} (2015) 026 [{\tt arXiv:1412.1919 [hep-th]]}. 
  
 \bibitem{Berman2015}
D.S. Berman and D.C.Thompson, \emph{Duality Symmetric String and M-theory}, Phys. Rept. {\bf 566} (2014) 1-60  [{\tt arXiv:1306.2643 [hep-th]}].

  
  \bibitem{Pezzella2015}
F. Pezzella,  \emph{Two Double String Theory Actions: non-covariance vs. covariance},  PoS CORFU2014 (2015) 158 [{\tt arXiv:1503.01709v2 [hep-th]}].

\bibitem{Pezzella2015a}
F. Pezzella, \emph{Some Aspects of the T-duality Symmetric String $\sigma$-model}, in Proceedings \emph{14th Marcel Grossmann Meeting on Recent Developments in Theoretical and Experimental General Relativity, Astrophysics and Relativistic Field Theories (MG14)} (in 4 volumes), \emph{Rome, Italy, July 12-18, 2015} vol. 4, pp. 4228-4233 {\tt [arXiv:1512.08825  [hep-th]]}.

\bibitem{Bandos2015}
I. Bandos, \emph{Superstring in Doubled Superspace}, Phys. Lett. {\bf B751} (2015) 408 [{\tt arXiv:1507.07779 [hep-th]}].

\bibitem{HZ}
C. Hull  and B. Zwiebach, \emph{Double Field Theory}, JHEP 0909 (2009) 099 {\tt [arXiv:0904.4664 [hep-th]]}.


\bibitem{hitchin1} N. J. Hitchin, \emph{Generalized Calabi-Yau manifolds}, Q. J. Math. 54 (2003) 281-308.
\bibitem{hitchin2} N. J. Hitchin, \emph{Lectures on Generalized Geometry}, {\tt [arXiv:1008.0973 [math.DG]]} .

\bibitem{gualtieri:tesi}
M. Gualtieri, \emph{Generalized Complex Geometry}, PhD Thesis,  {\tt [arXiv:math/0401221]}.

\bibitem{FRS}
L. Freidel, F.J. Rudolph, and D. Svoboda, \emph{A Unique Connection for Born Geometry}, LMU-ASC 37-18 [{\tt arXiv:1806.05992 [hep-th]]} .


\bibitem{Ossa1993}
X. de la Ossa and F. Quevedo, \emph{Duality Symmetries from non-Abelian isometries in String Theory},  Nucl. Phys. {\bf B403} (1993) 377 [{\tt hep-th/9210021}].

\bibitem{RV93}
M. Ro\v{c}ek and E. Verlinde, \emph{Duality, Quotients, and Currents},  Nucl. Phys. B373 (1992) 630 [{\tt arXiv:hep-th/9110053 }]

\bibitem{Klimcik1996a} 
C. Klim\v{c}ik  and P. \v{S}evera, \emph{Poisson-Lie T-Duality and Loop Groups of Drinfel'd Doubles}, Phys. Lett. {\bf B372}  (1996) 65 [{\tt  hep-th/9512040}].

\bibitem{Klimcik1996}
C. Klim\v{c}ik, \emph{Poisson-Lie T-Duality}, Nucl. Phys. Proc. Suppl. {\bf 46} (1996) 116 [{\tt hep-th/9509095}].

\bibitem{Drinfeld1987}
V.G. Drinfel'd, \emph{Quantum Groups}, Proceedings of the International Congress of Mathematicians (Berkley,  Caiif., 1986), American Mathematical Society, Providence USA (1987), pp. 798-820.

\bibitem{MPV2019}
V.E. Marotta, F. Pezzella, and  P. Vitale,
\emph{T-Dualities and Doubled Geometry of the Principal Chiral Model}, 
JHEP {\bf 1911} (2019) 123
{\tt [arXiv:1903.01243  [hep-th]]}

\bibitem{MPV2018} 
V.E. Marotta, F. Pezzella, and  P. Vitale,
  \emph{Doubling, T-Duality and Generalized Geometry: a Simple Model},
  JHEP {\bf 1808} (2018) 185 
  %doi:10.1007/JHEP08(2018)185
  {\tt [arXiv:1804.00744 [hep-th]]}
  
  \bibitem{BPV} F. Bascone,   F. Pezzella, and P. Vitale, \emph{Poisson-Lie T-Duality of WZW Model via Current Algebra Deformation},  [{\tt e-Print: 2004.12858 [hep-th]}] .
  
\bibitem{MO1} J. Maldacena and H. Ooguri, \emph{Strings in $AdS_{3}$ and $SL(2, R)$ WZW Model. Part I: The Spectrum}, J. Math.Phys. {\bf 42} (2001) 2929 {\tt [arXiv:hep-th/0001053]}.

\bibitem{MO2}  J. Maldacena and H. Ooguri, \emph{Strings in $AdS_{3}$ and $SL(2, R)$ WZW Model. Part II: Euclidean Black Hole }, J. Math. Phys. {\bf 42} (2001) 2961 {\tt[arXiv:hep-th/0005183]}.

\bibitem{MO3} J. Maldacena and H. Ooguri, \emph{Strings in $AdS_{3}$ and $SL(2, R)$ WZW Model. Part III: Correlation Functions}, Phys. Rev. {\bf D65} (2002) 106006 {\tt{arXiv:hep-th/0111180}]}.

\bibitem{gotz} 
  G.~Gotz, T.~Quella and V.~Schomerus,
  %``The WZNW model on PSU(1,1|2),''
  JHEP {\bf 0703} (2007) 003
  %doi:10.1088/1126-6708/2007/03/003
 {\tt  [hep-th/0610070]}.
  
  \bibitem{nappi}
  C.~R.~Nappi and E.~Witten,
  Phys.\ Rev.\ Lett.\  {\bf 71} (1993) 3751
  %doi:10.1103/PhysRevLett.71.3751
 {\tt  [hep-th/9310112]}.
  
\bibitem{kehagias}
  A.~A.~Kehagias and P.~Meessen,
  Phys.\ Lett.\ B {\bf 331} (1994) 77
  %doi:10.1016/0370-2693(94)90945-8
{\tt   [hep-th/9403041]}. 
  
     \bibitem{wittenbh}
  E.~Witten,
  %``On string theory and black holes,''
  Phys.\ Rev.\ D {\bf 44} (1991) 314.
  doi:10.1103/PhysRevD.44.314

 \bibitem{Sfetsos:1998} 
 K. Sfetsos, {\em Poisson-Lie T-duality and Supersymmetry}, Nucl. Phys. {\bf B} Proc. Suppl. {\bf 56B} (1997) 302, {\tt [hep-th/9611199]}; {\em Canonical equivalence of nonisometric sigma models and Poisson-Lie T-duality}, {\em Nucl. Phys. {\bf B517} (1998} 549 {\tt [hep-th/9710163]}; 
  {\em Poisson-Lie T duality beyond the classical level and the renormalization group},
  Phys.\ Lett.\ B {\bf 432} (1998) 365 
  %doi:10.1016/S0370-2693(98)00666-2
 {\tt [hep-th/9803019]}.
  %%CITATION = doi:10.1016/S0370-2693(98)00666-2;%% 
  
  \bibitem{Stern:1998} 
  A. Stern,
  \emph{Hamiltonian approach to Poisson Lie T - duality,}
  Phys.\ Lett.\ B {\bf 450} (1999) 141 
  %doi:10.1016/S0370-2693(99)00111-2
  {\tt [hep-th/9811256]}.
  %%CITATION = doi:10.1016/S0370-2693(99)00111-2;%%
  
  \bibitem{Stern:1999} 
  A. Stern,
\emph{T duality for coset models,}
  Nucl.\ Phys. {\bf B557} (1999) 459
  %doi:10.1016/S0550-3213(99)00397-1
  {\tt [hep-th/9903170]}.
  %%CITATION = doi:10.1016/S0550-3213(99)00397-1;%%
  
\bibitem{Falceto:2001} 
 F.  Falceto and K. Gawedzki,
 \emph{Boundary G/G theory and topological Poisson-Lie sigma model},
  Lett.\ Math.\ Phys.\  {\bf 59} (2002) 61 
 % doi:10.1023/A:1014477117077
  {\tt [hep-th/0108206]}.
  %%CITATION = doi:10.1023/A:1014477117077;%%
  
  \bibitem{Calvo:2003} 
  I. Calvo, F. Falceto and  D. Garcia-Alvarez,
\emph{Topological Poisson sigma models on Poisson lie groups},
  JHEP {\bf 0310} (2003) 033 
  %doi:10.1088/1126-6708/2003/10/033
  {\tt [hep-th/0307178]}.
  %%CITATION = doi:10.1088/1126-6708/2003/10/033;%%
  
  
  \bibitem{Bonechi:2003} 
 F.  Bonechi  and M. Zabzine,
 \emph{Poisson sigma model over group manifolds},
  J.\ Geom.\ Phys.\  {\bf 54}  (2005) 173
 % doi:10.1016/j.geomphys.2004.09.004
 {\tt [hep-th/0311213]}.
  %%CITATION = doi:10.1016/j.geomphys.2004.09.004;%%
  
 \bibitem{Sfetsos:2009} K. Sfetsos and K. Siampos,
  \emph{ Quantum equivalence in Poisson-Lie T-duality}, 
  JHEP {\bf 0906} (2009) 082 
  %doi:10.1088/1126-6708/2009/06/082
 {\tt [arXiv:0904.4248 [hep-th]]}.
  %%CITATION = doi:10.1088/1126-6708/2009/06/082;%%
  
 \bibitem{Severa:2017} 
  P. Severa,
  {\em On integrability of 2-dimensional $\sigma$-models of Poisson-Lie type},
  JHEP {\bf 1711} (2017) 015 
 % doi:10.1007/JHEP11(2017)015
 {\tt  [arXiv:1709.02213 [hep-th]]}.
  %%CITATION = doi:10.1007/JHEP11(2017)015;%%

  \bibitem{Hassler:2017} 
  F. Hassler,
  \emph{Poisson-Lie T-Duality in Double Field Theory},
{\tt  [arXiv:1707.08624 [hep-th]]}.
  %%CITATION = ARXIV:1707.08624;%%
  
  \bibitem{Jurco:2017} 
 B.  Jurco and J.  Vysoky,
  \emph{Poisson-Lie T-duality of string effective actions: A new approach to the dilaton puzzle,}
  J.\ Geom.\ Phys.\  {\bf 130} (2018) 1 .
  %doi:10.1016/j.geomphys.2018.03.019
  {\tt [arXiv:1708.04079 [hep-th]]}.
  %%CITATION = doi:10.1016/j.geomphys.2018.03.019;%%
  
  

 \bibitem{R89}
S. G. Rajeev, \emph{Non Abelian Bosonization without Wess-Zumino terms. 1. New current algebra}, Phys. Lett. {\bf B} 217 (1989) 123-128 .
\bibitem{R892}
S. G. Rajeev, 
  \emph{Nonabelian Bosonization Without Wess-Zumino Terms. 2.}, Aug. 1988,  UR-1088.
  
 \bibitem{Witten1983} 
 E. Witten,
  \emph{Nonabelian Bosonization in Two-Dimensions,}
  Commun.\ Math.\ Phys.\  {\bf 92} (1984) 455.
  %doi:10.1007/BF01215276
  
  \bibitem{RB1984} 
  G. Bhattacharya and S. Rajeev,
\emph{Boson - Fermion Equivalence in a Two-dimensional Anomalous Chiral Model,}
  Nucl.\ Phys.\ B {\bf 246} (1984) 157.
  %%CITATION = doi:10.1016/0550-3213(84)90119-6;%% 
  
 \bibitem{MSS92}
G. Marmo, A. Simoni, and A. Stern, \emph{Poisson-Lie group symmetries for the isotropic rotor}, Int. J. Mod. Phys. A 10, 99 (1995),  {\tt [arXiv:hep-th/9310145v1] }.
  \bibitem{RSV93}
S. G. Rajeev, G. Sparano, and P. Vitale, \emph{Alternative Canonical Formalism for the Wess-Zumino-Witten Model}, Int. J. Mod. Phys. A 9(31) 5469-5487 {\tt [arXiv:hep-th/9312178v1]}.

\bibitem{Borngeomf} 
L. Freidel, R.~G. Leigh  and D. Minic, \emph{Born Reciprocity in String Theory and the Nature of Spacetime,}
  Phys.\ Lett.\ B {\bf 730} (2014) 302
  %doi:10.1016/j.physletb.2014.01.067
 {\tt  [arXiv:1307.7080 [hep-th]]}. 
  %%CITATION = doi:10.1016/j.physletb.2014.01.067;%%
  
  \bibitem{marottaszabo}  V. E. Marotta and  R. J. Szabo,  \emph{Para-Hermitian Geometry, Dualities and Generalized Flux Backgrounds}, Fortschritte der Physik, [1800093] % DOI: 10.1002/prop.201800093, 
{\tt [arXiv:1810.03953 [hep-th]]}.

\bibitem{BMPV} F. Bascone, V. E. Marotta, F. Pezzella,   and P., \emph{T-Duality and Doubling of the Isotropic Rigid Rotator},  PoS CORFU2018 (2019) 123, {\tt [arXiv:1904.03727 [hep-th]]}.

\end{thebibliography}
\end{document}